\documentclass[11pt,a4paper]{article}
\pdfoutput=1 
\usepackage{jheppub}
\usepackage{todonotes}
\usepackage{amsmath}  
\usepackage{CJKutf8}
\usepackage{physics2}
\usephysicsmodule{ab,ab.braket} 
\usepackage{tcolorbox}
\usepackage{amsfonts}   
\usepackage{amsthm}    
\usepackage{mathtools} 
\usepackage{mathdots}

\usepackage{graphicx}
\usepackage{subcaption}
\usetikzlibrary{arrows.meta,shapes.misc,decorations.markings}

\tikzset{cross/.style={cross out, draw=black, minimum size=2*(#1-\pgflinewidth), inner sep=0pt, outer sep=0pt},
cross/.default={2pt}}
\definecolor{Pblue}{RGB}{0, 68, 105}
\definecolor{kblue}{RGB}{0,47,167}
\definecolor{usdGreen}{RGB}{11, 103, 55}

\RequirePackage{hyperref}
\hypersetup{
  colorlinks,citecolor= kblue,linkcolor= kblue,urlcolor=kblue
}
\definecolor{ultramarine}{RGB}{16,9,143}

\newcommand{\me}{\mathrm{e}}  
\newcommand{\mi}{\mathrm{i}} 
\newcommand{\dif}{\mathrm{d}} 
\newcommand{\Li}{\operatorname{Li}} 
\newcommand{\dbar}{{\mkern4mu\mathchar'26\mkern-11mu \mathrm{d}}}
\newcommand{\Res}{\operatorname{Res}}
\newcommand{\calM}{\mathcal{M}}
\newcommand{\calG}{\mathcal{G}}

\newcommand{\shortvdash}{%
  \mathrel{\tikz[baseline=-0.55ex,line width=0.50pt]{
    \draw (0,-0.65ex) -- (0,0.65ex);
    \draw (0,0) -- (0.30em,0ex);
  }}%
}
\newcommand{\shorttop}{%
  \mathrel{\tikz[baseline=-0.35em,line width=0.50pt]{
    \draw (-0.65ex,0) -- (0.65ex,0);
    \draw (0,0) -- (0.0em,-0.35em);
  }}%
}
\newcommand{\shortbot}{%
  \mathrel{\tikz[baseline=0.0em,line width=0.50pt]{
    \draw (-0.65ex,0) -- (0.65ex,0);
    \draw (0,0) -- (0.0em,0.36em);
  }}%
}

\renewcommand{\ln}{\log}

\preprint{BONN-TH/2026-17}

\title{Three-Reggeon exchange in $\mathcal{N}=4$ SYM to leading logarithmic accuracy}
\author[a]{Claude Duhr,}
\author[b]{Zhenjie Li \begin{CJK*}{UTF8}{gbsn}(李振杰)\end{CJK*}}
\author[c]{and Chi Zhang \begin{CJK*}{UTF8}{gbsn}(张驰)\end{CJK*}}

\affiliation[a]{Bethe Center for Theoretical Physics and Cluster of Excellence ``Color meets Flavor'', Universit\"{a}t Bonn, D-53115 Bonn, Germany}
\affiliation[b]{SLAC National Accelerator Laboratory, Stanford University, Stanford, CA 94309, USA}
\affiliation[c]{Bethe Center for Theoretical Physics, Universität Bonn, D-53115 Bonn, Germany}

\emailAdd{cduhr@uni-bonn.de}
\emailAdd{munuxilee@gmail.com}
\emailAdd{czhang@uni-bonn.de}

\abstract{We study eight-point amplitudes in the planar $\mathcal{N}=4$ Super Yang-Mills theory in multi-Regge kinematics in the Mandelstam region that receives contributions from both two- and three-Reggeon exchange. We use an effective field theory approach to compute the leading contribution from three-Reggeon exchange, and we explicitly evaluate the relevant diagrams up to four loops. We also propose a compact Fourier-Mellin representation for the contribution from two-Reggeon exchange in this region which involves the same ingredients as in other Mandelstam regions and can in principle be evaluated to any desired order in perturbation theory. To validate our proposal, we show that we can reproduce the multi-Regge limit of the known results for octagons up to three loops. We then combine the contributions from two- and three-Reggeon exchange to obtain novel results for octagons in MRK up to next-to-leading-logarithmic accuracy at four loops for the maximally helicity violating (MHV) configuration, and up to three loops for non-MHV contributions. We also discuss how our result can be extended to more particles, and we present the contribution from three-Reggeon exchange for three-loop MHV amplitudes with an arbitrary number of legs. In the appendices, we provide details on how to perform the analytic continuation and take the multi-Regge limit of the three-loop octagon, and explain how our results are connected to previous results in the literature.}

\makeatletter
\gdef\@fpheader{\phantom{M}}
\makeatother

\begin{document}
\maketitle
\flushbottom

\section{Introduction}

The $S$-matrix of the planar limit of the $\mathcal{N}=4$ Super Yang-Mills (SYM) theory is endowed with remarkable properties. In particular, it enjoys an infinite-dimensional Yangian symmetry~\cite{Drummond:2009fd}, generated by the ordinary and dual superconformal generators~\cite{Drummond:2006rz,Drummond:2008vq}. The dual conformal invariance is broken by the infrared (IR) singularities of the scattering amplitudes~\cite{Drummond:2007au}. Due to the universal nature of IR singularities, it is possible to define IR-finite and dual conformally invariant quantities. Specifically, the IR divergences are captured by the celebrated BDS ansatz~\cite{Anastasiou:2003kj,Bern:2005iz}. While it was originally conjectured to describe all maximally helicity violating (MHV) amplitudes to all orders in the coupling, the BDS ansatz needs to be corrected by a dual conformally invariant remainder function starting from six external particles~\cite{Drummond:2007au}. In addition to this enlarged symmetry, there are various dual descriptions of the $S$-matrix, for example via the celebrated AdS/CFT duality~\cite{Maldacena:1997re,Witten:1998qj,Alday:2007hr}, or via the description of scattering amplitudes in the near-collinear limit as excitations of a flux tube~\cite{Basso:2013vsa}. 

The aforementioned dualities and symmetries have led to novel approaches to compute scattering amplitudes in planar $\mathcal{N}=4$ SYM, including bootstrap approaches in which analytic results are obtained by constructing appropriate bases of functions whose coefficients are fixed by matching to appropriate boundary data. In particular, complete analytic results exist for MHV and next-to-MHV (NMHV) amplitudes for six particles up to eight loops~\cite{DelDuca:2009au,DelDuca:2010zg,Goncharov:2010jf,Dixon:2011pw,Dixon:2013eka,Dixon:2014iba,Dixon:2014voa,Dixon:2015iva,Caron-Huot:2020bkp,Caron-Huot:2019vjl,Dixon:2023kop}, for seven particles up to five loops~\cite{Golden:2014xqf,Dixon:2016nkn,Drummond:2017ssj,Dixon:2020cnr,He:2025tyv}, and for eight particles up to three loops~\cite{Caron-Huot:2011zgw,He:2019jee,He:2020vob,Li:2021bwg}. Despite this impressive progress, results for amplitudes in general kinematics with more loops and/or legs, or for other helicity configurations, are currently only available at two loops~\cite{Caron-Huot:2011zgw,Golden:2013lha,Kristensson:2021ani,Spiering:2024sea,He:2022ujv}, due to the complexity of the required computations.

In order to make progress and to gather data (which may also serve as boundary data in the bootstrap program), it is useful to turn to specific kinematic limits. One such limit is the multi-Regge limit, where the produced particles are strongly ordered in rapidity. In fact, the multi-Regge limit was one of the first places where the incompleteness of the BDS conjecture starting from six particles was pointed out~\cite{Bartels:2008ce}. Indeed, in Euclidean kinematics in the multi-Regge limit the BDS ansatz is valid for any number of external particles, and consequently the remainder function vanishes in Euclidean multi-Regge kinematics (MRK)~\cite{Bartels:2008ce,Brower:2008nm,Brower:2008ia,DelDuca:2008jg}. However, if the amplitude is analytically continued to a specific Mandelstam region where some of the produced particles have negative energy, the BDS ansatz in MRK develops an analytic structure that violates the Steinmann conditions~\cite{Steinmann,Steinmann2,Cahill:1973qp}, and this violation needs to be compensated by the dual conformally invariant remainder function.

It is possible to describe BDS-normalized amplitudes in MRK using effective field theory approaches as an exchange of bound states called \emph{Reggeons}, cf.,~e.g.,~refs.~\cite{Lipatov:1995pn,Antonov:2004hh}. In particular, in the Mandelstam region where all the centrally-produced particles have negative energy, the amplitude only receives contributions from two-Reggeon exchange. It is then possible to resum large logarithms in MRK to all orders in perturbation theory via the celebrated Balitsky-Fadin-Kuraev-Lipatov (BFKL) equation~\cite{Kuraev:1977fs,Balitsky:1978ic}. The amplitude factorizes in an auxiliary Fourier-Mellin space~\cite{Bartels:2009vkz,Bartels:2011ge}, and the ingredients in Fourier-Mellin space can be described to all orders in the coupling using integrability and an analytic continuation from the flux-tube description in the near-collinear limit~\cite{Basso:2014pla,DelDuca:2019tur}. While this connection to integrability is conjectural, it is consistent with explicit fixed-order results~\cite{Bartels:2009vkz,Lipatov:2010ad,Fadin:2011we,Bartels:2011ge,Dixon:2012yy,Dixon:2014voa,DelDuca:2018hrv}. In addition, there are powerful techniques to evaluate the Fourier-Mellin integrals in terms of single-valued polylogarithms~\cite{BrownSVHPLs,brownSV,Brown:2013gia} to very high orders in the coupling and for many external particles~\cite{Lipatov:2010ad,Prygarin:2011gd,Fadin:2011we,Bartels:2011ge,Dixon:2012yy,Pennington:2012zj,Broedel:2015nfp,Bargheer:2015djt,DelDuca:2016lad,Broedel:2016kls,DelDuca:2018raq,DelDuca:2018hrv,Marzucca:2018ydt,DelDuca:2019tur,Bargheer:2019lic,Baune:2023uut}. Results at strong coupling are also available~\cite{Bartels:2012gq,Bartels:2014ppa,Bartels:2014mka}.

Starting from eight points, Mandelstam regions where not all the centrally-produced particles have negative energy appear, and they are not as well understood. From an effective field theory perspective, this can be traced back to the fact that these amplitudes receive contributions from an exchange of three (or more) Reggeons~\cite{Lipatov:2009nt,Bartels:2011nz,Caron-Huot:2013fea,Bartels:2020twc}. The simplest instance where three-Reggeon exchange contributes is the eight-point amplitude in the so-called \emph{zigzag region}. Currently, there are no results for scattering amplitudes in MRK in Mandelstam regions where three-Reggeon exchange gives a nontrivial contribution.

The main goal of this paper is to take first steps toward analytic results for scattering amplitudes in MRK in Mandelstam regions that receive contributions from three-Reggeon exchange. We focus on three-Reggeon exchange at leading logarithmic accuracy (LLA). We use an effective field theory approach and we obtain a simple set of Feynman rules that allow us to extract the contribution from multi-Reggeon exchange at LLA. Three-Reggeon exchange at LLA, however, only contributes to the full amplitude in MRK at next-to-LLA (NLLA). Hence, our three-Reggeon contributions need to be consistently combined with the contributions from two-Reggeon exchange at NLLA. One of our main results is a compact Fourier-Mellin representation for two-Reggeon exchange in these regions, valid to all orders in the coupling. As a cross-check of our results, we show that our proposal has the correct behavior in soft limits and that it reproduces the multi-Regge limits of the known BDS-normalized eight-point amplitudes from refs.~\cite{Caron-Huot:2011zgw,Bargheer:2015djt,DelDuca:2018raq,He:2019jee,Li:2021bwg}. We then combine our results for two- and three-Reggeon exchange to obtain novel results in MRK at NLLA in the zigzag region for the four-loop eight-point MHV amplitude, as well as for non-MHV amplitudes at three loops. We also present a compact result for the three-Reggeon contribution to all three-loop MHV amplitudes in MRK in such a Mandelstam region.

Our paper is organized as follows: In section~\ref{sec:review} we review the general background on scattering amplitudes in $\mathcal{N}=4$ SYM in MRK, and in section~\ref{sec:twoReggeon} we recall the effective field theory approach to MRK. In section~\ref{sec:three-reggeon} we present our main result, namely we describe the structure of the eight-point amplitude in MRK in the zigzag region to NLLA, including contributions from both two- and three-Reggeon exchange. We also show that we reproduce the known results for the MHV amplitude up to three loops, and we present new results at higher loops and for non-MHV configurations. In section~\ref{sec:conclusion} we draw our conclusions. We also include an appendix where we discuss how to perform the analytic continuation and how to take the multi-Regge limit for the three-loop octagon.

\section{Review of amplitudes and multi-Regge kinematics in \texorpdfstring{$\mathcal{N}=4$}{N=4} SYM}
\label{sec:review}

In this section, we review some basic properties of scattering amplitudes and multi-Regge kinematics in the planar $\mathcal{N}=4$ SYM theory. We also establish the notation and conventions used throughout this paper.

\subsection{Scattering amplitudes in planar \texorpdfstring{$\mathcal{N}=4$}{N=4} SYM}

Scattering amplitudes in planar $\mathcal{N}=4$ SYM are free of ultraviolet divergences and are highly constrained by the extensive symmetries of the theory. At tree level, these constraints are organized into an infinite-dimensional Yangian symmetry generated by the superconformal and dual superconformal symmetries (for a review, see ref.~\cite{Drummond:2010km}). Although both symmetries are broken by the infrared divergences of loop amplitudes, the infrared (IR) structure of the amplitudes remains remarkably simple and is well understood. More precisely, it is encoded into the so-called BDS ansatz~\cite{Bern:2005iz}, which is usually formulated in dimensional regularization as
\begin{equation} \label{eq:BDS}
  A^{\text{BDS}}_{N}=A_{N,\text{MHV}}^{(0)} \exp\ab[\,\sum_{L=1}^{\infty}g^{2L}\ab(f^{(L)}(\epsilon)M_{N}^{(1)}(L\epsilon)+C^{(L)})] \:,
\end{equation}
where $g^2\equiv N\!_{c}\,g_{\text{YM}}^2/(16\pi^2)$ is the coupling constant (with $N\!_c$ the number of colors) and the superscript $(L)$ denotes the loop order in the perturbative expansion, $C^{(L)}$ are constants, and $f^{(L)}(\epsilon):=f_{0}^{(L)}+\epsilon f_{1}^{(L)}+\epsilon^2 f_{2}^{(L)}$ is quadratic in $\epsilon$. The coefficients $f_{0}^{(L)}$ are determined by the cusp anomalous dimension,
\begin{equation}
  f_{0}:=\sum_{L=1}^{\infty}g^{2L}f_{0}^{(L)}=\gamma_{K}/8=g^2-2\zeta_{2}\,g^4+22\zeta_{4}\,g^6+\cdots \:,
\end{equation}
which is known to all orders from integrability~\cite{Beisert:2006ez}. The remaining coefficients $f_{1}^{(L)}$, $f_{2}^{(L)}$, and $C^{(L)}$ are known up to certain loop orders~\cite{Bern:2005iz,Cachazo:2007ad,Dixon:2017nat,Agarwal:2021zft}. The BDS ansatz is essentially given by the exponentiation of the one-loop maximally helicity violating (MHV) amplitude. Its kinematic dependence is entirely determined by the tree-level MHV amplitude $A_{N,\mathrm{MHV}}^{(0)}$ and the normalized one-loop amplitude $M_{N}^{(1)}(\epsilon):=A_{N,\text{MHV}}^{(1)}/A_{N,\text{MHV}}^{(0)}$.

We are now in a position to introduce our main object of interest, the BDS-normalized amplitude for gluon scattering,
\begin{equation}
  R_{N}^{h_1\cdots h_N} := \frac{A_{N}^{h_1\cdots h_N}/A_{N}^{(0),h_1\cdots h_N}}{A_{N}^{\text{BDS}}/A_{N,\text{MHV}}^{(0)}} \:,
\end{equation}
where $A_{N}^{h_1\cdots h_N}$ denotes the $N$-point amplitude, and $h_{i}=\oplus\ (\ominus)$ denotes the positive (negative) helicity of particle $i$ (considered as outgoing). Here, for later convenience, we have also normalized the numerator and denominator by their respective tree-level counterparts.
This quantity is finite and, more importantly, dual conformally invariant. It is then convenient to introduce the \emph{dual coordinates} $x_{i}$, defined via
\begin{equation} \label{eq:def_x}
   x_{ij}^2\equiv(x_{i}-x_{j})^2=(p_{i+1}+p_{i+2}+\ldots+p_{j})^2 \:.
\end{equation}
The dual conformal invariance renders the BDS-normalized amplitude a function of cross-ratios of $x_{i}$ only. 
In this paper, we mainly focus on the MHV case, where, without loss of generality, we choose particles 1 and 2 to be of negative helicity. We therefore denote the MHV BDS-normalized amplitude by $R_{N}=R_{N}^{\ominus\ominus\oplus\cdots\oplus}$ unless otherwise indicated. It then follows from the definition of the BDS ansatz in eq.~\eqref{eq:BDS} that the perturbative expansion of $R_{N}$ takes the form
\begin{equation}
  R_{N}=1+\sum_{L=2}^{\infty} g^{2L} R_{N}^{(L)} \:.
\end{equation}

\subsection{Multi-Regge kinematics}

\begin{figure}[t]
  \centering

  \begin{subfigure}[b]{0.45\textwidth}
    \centering
    \begin{tikzpicture}
      \draw[line width=0.3mm] (0,-1.5) -- (0,1.5);
      \draw[line width=0.3mm] (0,1.5) -- +(5:1.5)node[right]{$p_3$} (0,1.5) -- +(180-5:1.5)node[left]{$p_2$};
      \draw[line width=0.3mm] (0,-1.5) -- +(-5:1.5)node[right]{$p_N$} (0,-1.5) -- +(-175:1.5)node[left]{$p_1$};
      \draw[line width=0.3mm] (0,0.75) -- (1.4,0.75)node[right]{$p_4$};
       \draw[line width=0.3mm] (0,-0.75) -- (1.4,-0.75)node[right]{$p_{N-1}$};
       \node at (-0.75,0) {$x_{1}$};
        \node at (0,1.9) {$x_{2}$};
        \node at (0,-1.9) {$x_{N}$};
         \node[right] at (0.70,-1.2) {$x_{N-1}$};
             \node[right] at (0.70,1.2) {$x_{3}$};
      \foreach \y in {-1,0,1}
      \fill (0.45,\y*0.3) circle (1.2pt);
    \end{tikzpicture}
    \caption{}
    \label{fig:mrk-a}
  \end{subfigure}
  \hfill
  \begin{subfigure}[b]{0.45\textwidth}
    \centering
    \begin{tikzpicture}
       \draw[line width=0.3mm] (0,-1.5) -- (0,1.5);
      \draw[line width=0.3mm] (0,1.5) -- +(5:1.5)node[right]{$\mathbf{p}_3$} (0,1.5) -- +(180-5:1.5)node[left]{$\mathbf{0}$};
      \draw[line width=0.3mm] (0,-1.5) -- +(-5:1.5)node[right]{$\mathbf{p}_N$} (0,-1.5) -- +(-175:1.5)node[left]{$\mathbf{0}$};
      \draw[line width=0.3mm] (0,0.75) -- (1.4,0.75)node[right]{$\mathbf{p}_4$};
       \draw[line width=0.3mm] (0,-0.75) -- (1.4,-0.75)node[right]{$\mathbf{p}_{N-1}$};
       \node at (-0.75,0) {$\mathbf{x}_{1}$};
        \node at (0,1.9) {$\mathbf{x}_{1}$};
        \node at (0,-1.9) {$\mathbf{x}_{1}$};
         \node[right] at (0.70,-1.2) {$\mathbf{x}_{N-2}$};
             \node[right] at (0.70,1.2) {$\mathbf{x}_{2}$};
      \foreach \y in {-1,0,1}
      \fill (0.45,\y*0.3) circle (1.2pt);
    \end{tikzpicture}
    \caption{}
    \label{fig:mrk-b}
  \end{subfigure}
  \caption{An illustration of the multi-Regge kinematic configuration.}
  \label{fig:mrk}
\end{figure}
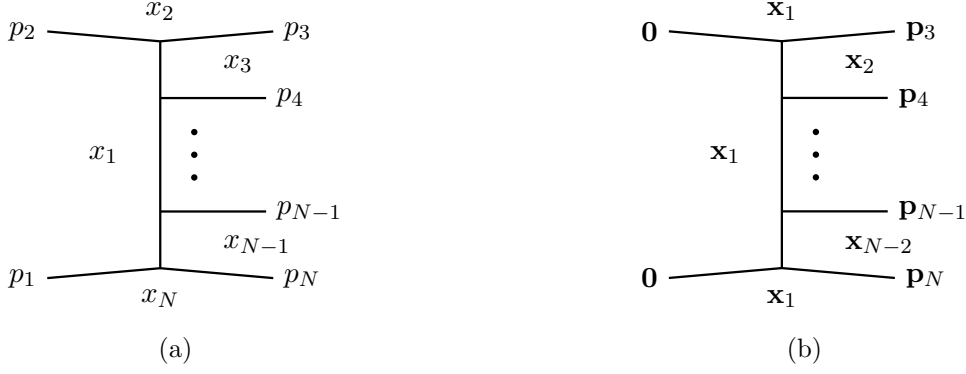

Multi-Regge kinematics (MRK) are designed to isolate the $t$-channel dynamics of high-energy scattering processes. This configuration is illustrated in figure~\ref{fig:mrk-a}, where particles $1$ and $2$ are chosen as the incoming particles, particles $3$ and $N$ are outgoing, and the remaining ones are associated with the $t$-channel exchanges.

We choose a reference frame in which the incoming particles collide along the $z$-axis. The multi-Regge limit then requires a strong hierarchy among the rapidities of the remaining particles, while their transverse momenta remain of the same order. 
In light-cone coordinates, this kinematic configuration can be expressed as
\begin{gather}
\notag p_{1}=(0,p_{1}^{-},\mathbf{0}),\qquad
p_{2}=(p_{2}^{+},0,\mathbf{0}) \:, \\
 \lvert p_{3}^{+}\rvert \gg \lvert p_{4}^{+}\rvert \gg \cdots \gg \lvert p_{N}^{+}\rvert \:,  \hspace{1.4em}
 \lvert p_{3}^{-}\rvert \ll \lvert p_{4}^{-}\rvert \ll \cdots \ll \lvert p_{N}^{-}\rvert \:, \label{eq:p_hierarchy} \\
\notag \lvert \mathbf{p}_{3} \rvert \simeq \lvert \mathbf{p}_{4} \rvert \simeq \cdots \simeq \lvert \mathbf{p}_{N} \rvert \:,
\end{gather}
where $p_i^{\pm}:=p_i^t\pm p_i^z$, and $\mathbf{p}_i:=p_i^x+\mi\, p_i^y$ denotes the complexified transverse momentum. We denote the transverse component of $x_{i}$ by 
$\mathbf{x}_{i-1}$ for $3\leq i \leq N-1$, and we note that in our frame $\mathbf{p}_{1}=\mathbf{p}_2=\mathbf{0}$ (see figure~\ref{fig:mrk-b}). 
The multi-Regge limit implies the following hierarchy among the Lorentz invariants:
\begin{equation}
    |x_{2,N}^2| \gg |x_{2,N-1}^{2}|,|x_{3,N}^{2}| \gg\cdots\gg |x_{24}^{2}|,\cdots,|x_{N-2,N}^{2}| \gg |x_{13}^{2}|,\cdots,|x_{1,N-1}^{2}| \:.
\end{equation}
For later convenience, we introduce the abbreviations
\begin{equation}
  t_{i}:=x_{1,i+2}^2\simeq -|\mathbf{q}_{i}|^2,\quad k_{i}:=p_{i+3} \quad \text{and} \quad s_{i}:=x_{i+1,i+3}^2\,,
\end{equation}
for $1\leq i\leq N-3$, where $\mathbf{q}_{i}$ denotes the transverse component of the corresponding $t$-channel momentum transfer, and ``$\simeq$'' denotes equality in the multi-Regge limit, i.e., up to terms that are power-suppressed in the limit.

The hierarchy in the multi-Regge limit from eq.\ \eqref{eq:p_hierarchy} can be naturally parametrized by the ratios
$\delta_i:=p_{i+4}^{+}/p_{i+3}^{+} = (p_{i+3}^{-}/p_{i+4}^{-}) |\mathbf{p}_{i+4}/\mathbf{p}_{i+3}|^2$ with $0\leq i\leq N-4$. To leading power, the logarithms of these ratios have a physical interpretation as the rapidity gap between adjacent particles,
\begin{equation}  \label{eq:rap_to_s}
\eta_{i+3,i+4}:= \frac{1}{2} \log\ab|\frac{p_{i+3}^{+}p_{i+4}^{-}}{p_{i+3}^{-}p_{i+4}^{+}}|\simeq -\log \ab|
\frac{\delta_{i}\,\mathbf{p}_{i+3}}{\mathbf{p}_{i+4}}|
\simeq  \log \frac{|s_{i+1}|}{|\mathbf{p}_{i+3}\mathbf{p}_{i+4}|} \:.
\end{equation}  
In a general gauge theory, perturbative amplitudes at a fixed loop order are polynomials in these large logarithms  $\eta_{i+3,i+4}$ in the multi-Regge limit. However, the objects of interest in planar $\mathcal{N}=4$ SYM theory are the BDS-normalized amplitudes, which are dual conformally invariant. It is therefore convenient to introduce a dual-conformal completion of $\delta_i$ for $1\leq i \leq N-5$,
\begin{equation}
  \tau_{i}:= \sqrt{\frac{x^{2}_{N,i+3}x_{1,i+2}^2 x_{1,i+4}^2 x_{2,i+3}^2}{x^{2}_{N,i+2}x^{4}_{1,i+3}x^2_{2,i+4}}}
  \simeq |\delta_{i}|\, \frac{|\mathbf{q}_{i-1}\mathbf{q}_{i+1}\mathbf{p}_{i+3}|}{|\mathbf{q}_{i}|^2|\mathbf{p}_{i+4}|}\simeq \frac{|\mathbf{q}_{i-1}\mathbf{q}_{i+1}\mathbf{p}_{i+3}\mathbf{p}_{i+4}|}{|\mathbf{q}_{i}|^2|s_{i+1}|} \:.
\end{equation}

To fully specify the multi-Regge limit, one must also distinguish between different kinematic regions, often referred to as \emph{Mandelstam regions} in the literature~\cite{Lipatov:2010ad}.
This can be done by assigning definite signs $\varrho_i:=\operatorname{sign}(E_i)$ to the energies of particles $3,\ldots,N$.\footnote{Note that the signs of the energies of particles $1$ and $2$ are fixed to be opposite to those of particles $N$ and $3$, respectively, by momentum conservation and the hierarchy of scales in eq.~\eqref{eq:p_hierarchy}.} Each Mandelstam region can then be labeled by a sequence of signs $\varrho\equiv(\varrho_i)_{3\leq i\leq N}$. Note that, as a consequence of the hierarchy in eq.~\eqref{eq:p_hierarchy}, the consecutive Mandelstam variables in multi-Regge kinematics have definite signs in the Mandelstam region $\varrho$,
\begin{equation}
    \operatorname{sign}(x_{ij}^2)=
    \begin{cases}
        \varrho_{i+1}\varrho_{j}, &\quad  2\leq i <i+1<j \leq N\,, \\
        -1, &\quad 1=i<2<j<N\,.
    \end{cases}
\end{equation}
In the remainder of this paper, we will mainly deal with two classes of Mandelstam regions. For both classes, we adopt the notation introduced in ref.~\cite{Bargheer:2015djt}. The first class consists of regions with two negative entries, which we denote by 
\begin{equation}
\{k,l\}
=
(+,\cdots,+,\overset{k}{-},+,\cdots,+,
\overset{l}{-},+,\cdots,+)\, .
\end{equation}
The second class consists of regions with a consecutive block of negative entries, which we denote by
\begin{equation}
\label{eq:region_[k,l]}
[k,l]
=
(+,\cdots,+,\overset{k}{-},\cdots,
\overset{l}{-},+,\cdots,+)\, .
\end{equation}

In the following, it will be useful to denote by $\mathcal{M}\ab(A_N)$ the leading power in the multi-Regge limit of the $N$-point gluon scattering amplitude $A_N$, i.e., we have $A_N=\mathcal{M}\ab(A_N)+\ldots$, where the dots indicate terms that are power-suppressed in the multi-Regge limit.
We denote the BDS-normalized amplitude in the multi-Regge limit in the Mandelstam region $\varrho$ by
\begin{equation} \label{eq:nonMHVamp_MRK}
  \mathcal{R}_{\varrho,N}^{h_4\cdots h_{N-1}}:= \mathcal{M}\big(R_{\varrho,N}^{\ominus\ominus\oplus h_4\cdots h_{N-1}\oplus}\big)\,,
\end{equation} 
where $R_{\varrho,N}^{h_1\cdots h_{N}}$ denotes the analytic continuation of the amplitude $R_{N}^{h_1\cdots h_{N}}$ to the Mandelstam region $\varrho$. Note that, since helicity is conserved along the beam direction in the multi-Regge limit (that is, $h_3=-h_2$ and $h_N=-h_1$), and $h_{1}$ and $h_{2}$ are conventionally taken to be of negative helicity, we simply use $h_{4}\cdots h_{N-1}$ to denote the helicity configuration of the amplitude, as in eq.~\eqref{eq:nonMHVamp_MRK}. For MHV amplitudes, we also simply write $\mathcal{R}_{\varrho,N} = \mathcal{R}_{\varrho,N}^{\oplus\cdots \oplus}$.

The $L$-loop BDS-normalized MHV amplitude $\mathcal{R}_N^{(L)}$
is a polynomial in the logarithms $\log\tau_i$, and the MRK amplitude is graded by the total powers of $\log\tau_i$. The highest total power of $\log\tau_i$ is the LLA, the next-to-highest power is the NLLA, and so on. The coefficients of products of $\log\tau_i$ are functions of the transverse momenta only. Moreover, they are dual conformally invariant in the transverse momentum space, i.e., they are functions of cross ratios of the points $\mathbf{x}_i$. Two parametrizations of the transverse kinematic space modulo the residual dual conformal symmetry $\operatorname{SL}(2,\mathbb{C})$ have proven useful~\cite{DelDuca:2016lad} and will be used in this paper.
The first is given by the \emph{simplicial MRK coordinates}, defined by fixing the three points $(\mathbf{x}_{1},\mathbf{x}_{2},\mathbf{x}_{N-2})$ to $(1,0,\infty)$ and setting
\begin{equation}
\rho_i := \mathbf{x}_{i+2}\,, \qquad  1\leq i \leq N-5\,.
\end{equation}
The second is given by transverse cross ratios, also known as \emph{Fourier-Mellin coordinates},
\begin{equation}
z_i :=
\frac{
(\mathbf{x}_{1}-\mathbf{x}_{i+3})(\mathbf{x}_{i+2}-\mathbf{x}_{i+1})
}{
(\mathbf{x}_{1}-\mathbf{x}_{i+1})(\mathbf{x}_{i+2}-\mathbf{x}_{i+3})
}\,, \qquad 1\leq i \leq N-5 \,.
\end{equation}
We denote the complex conjugates of the $\mathbf{x}$-coordinates by $\mathbf{x}_{i}^{\ast}$, while those of the $\rho$- and $z$-coordinates are denoted by $\bar{\rho}_{i}$ and $\bar{z}_{i}$, respectively.

\subsection{The BDS ansatz in the multi-Regge limit and the BDS phase} \label{sec:BDS}

Scattering amplitudes in planar $\mathcal{N}=4$ SYM have been extensively studied. In particular, it is known that the BDS-normalized amplitudes become trivial in MRK in the Euclidean region where all consecutive Mandelstam invariants are negative. In other Mandelstam regions $\varrho$, the BDS-normalized amplitude may not be trivial, and its precise analytic form depends on $\varrho$. In order to understand the structure of the BDS-normalized amplitude in other Mandelstam regions, it is important to first understand the multi-Regge limit of the BDS ansatz. This has been studied in detail in refs.~\cite{DelDuca:2008pj,Bartels:2008ce,Brower:2008nm,DelDuca:2008jg,Brower:2008ia,DelDuca:2009ae,Bartels:2013jna,Bartels:2020twc}; here we merely quote the final result,

\begin{equation} \label{eq:BDSfactorInMRK}
    \mathcal{M}\ab(A_{\text{BDS}}^{\varrho}) = C^{\varrho}\:V_{\shorttop}(t_{1})
    \ab(\prod_{i=1}^{N-4}\ab(\frac{-s_{i}}{\mu^2})^{\omega_{g}(t_{i})} V^{\varrho}_{\shortvdash}\ab(\mathbf{q}_{i},\mathbf{q}_{i+1}) ) \ab(\frac{-s_{N-3}}{\mu^2})^{\omega_{g}(t_{N-3})} V_{\shortbot}(t_{N-3})\:,
\end{equation}
where $\mu^2$ is some IR regulator\footnote{The dependence on the dimensional regulator $\epsilon$ can be absorbed into the definition of $\mu^2$ and will be omitted below, see ref.~\cite{Bartels:2013jna} for more details.} and we drop the trivial tree-level factor for simplicity. The factor $C^{\varrho}$ is a pure phase which arises from the analytic continuation of dilogarithms in the BDS ansatz, and hence it first appears at six points.

Apart from the factor $C^{\varrho}$, eq.~\eqref{eq:BDSfactorInMRK} exhibits the typical factorization behavior in the multi-Regge limit. The factor $(-s_{i}/\mu^2)^{\omega_{g}(t_{i})}$ represents the propagator in the $t$-channel, with the gluon Regge trajectory given by
\begin{equation} \label{eq: gluon_regge_trajectory}
    \omega_{g}(t_{i})=-\frac{\gamma_{K}}{4}\ln \frac{-t_{i}}{\mu^{2}}\simeq-2\tilde{a}\ln\frac{|\mathbf{x}_{1,i+1}|^{2}}{\mu^{2}} \:,
\end{equation}
where we have introduced the abbreviation $\tilde{a}:=\gamma_{K}/8$ for later convenience. The factors $V_{\shorttop}$, $V_{\shortbot}$, and $V_{\shortvdash}$ denote the vertex factors associated with the top, bottom, and middle vertices, respectively. Their explicit expressions are of no interest here and can be found in refs.~\cite{Bartels:2008ce,Bartels:2013jna}; however, it is worth recording the dependence of $V_{\shortvdash}^{\varrho}$ on the Mandelstam region $\varrho$:
\begin{equation}
    \frac{V^{\varrho}_{\shortvdash}\ab(\mathbf{q}_{i},\mathbf{q}_{i+1})}{|V^{\varrho}_{\shortvdash}\ab(\mathbf{q}_{i},\mathbf{q}_{i+1})|} = \begin{cases}
        \me^{-\mi\pi \omega_{g}(\kappa_{i})/2}\:,
        & \text{if } \{\varrho_{i+2},\varrho_{i+3},\varrho_{i+4}\}=\{+,-,+\} \text{ or } \{-,+,-\}\:, \\
        \me^{+\mi\pi \omega_{g}(\kappa_{i})/2}\:, &\text{otherwise}\:,
    \end{cases}
\end{equation}
where
\begin{equation} \label{eq: one_loop_vertex_correction}
    \omega_{g}(\kappa_{i})=-\frac{\gamma_{K}}{8}\ln\frac{|\mathbf{q}_{i}|^{2}|\mathbf{q}_{i+1}|^{2}}{|\mathbf{q}_{i}-\mathbf{q}_{i+1}|^{2}\mu^{2}}=-\tilde{a}\ln\frac{|\mathbf{x}_{1,i+1}|^{2}|\mathbf{x}_{1,i+2}|^{2}}{|\mathbf{x}_{i+1,i+2}|^{2}\mu^{2}} \:.
\end{equation}
Note that the factor $(-s_i)^{\omega_g(t_i)}$ in eq.~\eqref{eq:BDSfactorInMRK} also yields a phase when $s_i>0$.

The naive factorization in eq.~\eqref{eq:BDSfactorInMRK} can be spoiled by the factor $C^{\varrho}$, which, starting at six points, may introduce kinematic dependence beyond $|\mathbf{q}_{i}|^2$ and $|\mathbf{k}_{i}|^2$, such as $|\mathbf{x}_{24}|^2$. Indeed, as found in ref.~\cite{Bartels:2008ce}, the multi-Regge limit of the six-point BDS ansatz in the region $[4,5]$ gives 
\begin{equation}
   C^{[4,5]}=\exp\ab( \mi\pi \frac{\gamma_{K}}{4}\ln\frac{|\mathbf{x}_{12}|^2|\mathbf{x}_{14}|^2}{|\mathbf{x}_{24}|^2\mu^2}) \:. \label{eq:sixpointBDSphase}
\end{equation}
At two-loop order, this factor (together with other factors in eq.~\eqref{eq:BDSfactorInMRK}) produces terms such as $\log|\mathbf{x}_{13}|^2\log|\mathbf{x}_{24}|^2$, which violate the Steinmann conditions~\cite{Steinmann,Steinmann2} and render the six-point BDS ansatz unphysical.
Therefore, the BDS ansatz is incomplete and must be supplemented with a remainder function, which is finite and dual conformally invariant and compensates for the violation of the Steinmann conditions by the BDS ansatz (see ref.~\cite{Lipatov:2010ad} for an explicit two-loop computation). As a consequence, the BDS-normalized amplitude is non-trivial. 

It is useful to isolate the unphysical pieces in a finite, dual-conformally invariant way. For example, the factor $C^{[4,5]}$ can be decomposed as
\begin{equation} \label{eq:remove_tri}
    \log C^{[4,5]} = \mi\pi \frac{\gamma_{K}}{8}\log\frac{|\mathbf{x}_{12}\mathbf{x}_{23}\mathbf{x}_{34}\mathbf{x}_{41}|^2}{|\mathbf{x}_{13}\mathbf{x}_{24}|^4} - \mi\pi\omega_{g}(\kappa_{1})-\mi\pi\omega_{g}(\kappa_{2}) \:,
\end{equation}
where the last two terms can be absorbed into $V_{\shortvdash}(\mathbf{q}_{i},\mathbf{q}_{i+1})$. More generally, we can express the multi-Regge limit of the BDS ansatz in a region $\varrho$ as 
\begin{equation} \label{eq:BDSinMRK}
    \calM(A_{\text{BDS}}^{\varrho})= \lvert \calM(A_{\text{BDS}})\rvert\exp(2\pi\mi\varphi^{\varrho}_{N})\exp(\pi\mi\tilde{a}\delta_{N}^{\varrho}),
\end{equation}
where $\varphi^{\varrho}_{N}$ consists only of $\omega_{g}(t_{i})$ and $\omega_{g}(\kappa_{i})$, and hence preserves the Steinmann conditions. The so-called \emph{BDS phase} $\delta_{N}^{\varrho}$ is a dual-conformally invariant completion of $C^{\varrho}$ and is responsible for the violation of the Steinmann conditions.

As the complete amplitude fulfills the Steinmann conditions, but the BDS ansatz does not, the BDS-normalized amplitude must also violate the Steinmann conditions. Therefore, the object of interest is the combination $\mathcal{R}_{\varrho,N}\me^{\pi\mi\tilde{a}\delta_{N}^{\varrho}}$. For one thing, it is finite and dual-conformally invariant. For another, it has the correct physical analytic structure, and hence may admit an interpretation as arising from an effective field theory.

It should be noted that the separation of $\me^{2\pi\mi\varphi^{\varrho}_{N}}$ and $\me^{\pi\mi\tilde{a}\delta^{\varrho}_{N}}$ in eq.\ \eqref{eq:BDSinMRK} is not unique.\footnote{For example, in ref.~\cite{Bartels:2013jna}, the BDS phase $\delta^{\{4,6\}}_{7}$ for the heptagon is chosen to be
\begin{equation*}
\delta^{\{4,6\}}_{7}=2\log\frac{|\mathbf{x}_{12}\mathbf{x}_{15}||\mathbf{x}_{24}\mathbf{x}_{35}|^2}{|\mathbf{x}_{25}|^2 |\mathbf{x}_{23}\mathbf{x}_{45}\mathbf{x}_{13}\mathbf{x}_{14}|} \:,
\end{equation*}
which is different from our choice in eq. \eqref{eq:BDSphase_choice}.
} A convenient choice for $\delta_{N}^{\varrho}$ is 
\begin{equation} \label{eq:BDSphase_choice}
    \delta^{\varrho}_{N} = \frac{1}{4} \sum_{i=2}^{N-4}\sum_{j=i+2}^{N-2}
    (\varrho_{i+1}-\varrho_{i+2})(\varrho_{j+1}-\varrho_{j+2})\log \biggl\lvert 
        \frac{\mathbf{x}_{i,j}^{2}\mathbf{x}_{1,i+1}\mathbf{x}_{j-1,1}}{\mathbf{x}_{1i}\mathbf{x}_{i,i+1}\mathbf{x}_{j-1,j}\mathbf{x}_{j,1}}
        \biggr\rvert ^{2} \:,
\end{equation}
which satisfies
\begin{equation} \label{eq: BDSphase_relation}
    \delta_{N}^{\{k,l\}}=\delta_{N}^{[k,l]}-\delta_{N}^{[k+1,l]}-\delta_{N}^{[k,l-1]}+\delta_{N}^{[k+1,l-1]}.
\end{equation}
This choice will be motivated in the next section.

\subsection{The BDS-normalized amplitude in the Mandelstam regions \texorpdfstring{$[k,l]$}{[k,l]}}

So far, mostly the Mandelstam regions of the form $[k,l]$ in eq.~\eqref{eq:region_[k,l]} have been studied in the literature, and for those regions there is a conjectural answer for the BDS-normalized amplitudes to all orders in perturbation theory (and possibly even at finite coupling).
 The key fact here is that the combination $\me^{\mi\pi\tilde{a}\delta^{\varrho}_{N}}\mathcal{R}_{\varrho,N}$ takes a factorized form in an auxiliary Fourier-Mellin space. More precisely, we have 
 \begin{align}\label{eq:2Reggeon_Ex}
 \mathcal{W}^{h_4\cdots h_{N-1}}_{[4,N-1]}&:=\frac{\mathcal{R}_{[4,N-1],N}^{h_4\cdots h_{N-1}}\,\me^{\pi\mi\tilde{a}\delta_{N}^{[4,N-1]}}}{2\pi\mi} \\
 \nonumber&\phantom{:}=g^2 \prod_{r=1}^{N-5}\ab[\sum_{n_r}\ab(\frac{z_{r}}{\bar{z}_{r}})^{n_{r}/2}\int_{\mathcal{C}}\frac{\dif \nu_{r}}{2\pi} \frac{1}{(-\tau_{r}+\mi 0)^{\omega_r}}]
  \chi^{h_4}_{1}C^{h_5}_{12}\cdots C^{h_{N-2}}_{N-6,N-5}\chi^{-h_{N-1}}_{N-5} \:, 
\end{align}
where we considered without loss of generality the Mandelstam region $[k,l] = [4,N-1]$. The three ingredients $\chi_i^{h} := \chi^{h}(\nu_i,n_i)$, $\omega_r := \omega(\nu_r,n_r)$ and $C^{h}_{ij} = C^{h}(\nu_i,n_i,\nu_j,n_j)$, with $h=\oplus/\ominus$, are the so-called \emph{impact factors}, \emph{BFKL eigenvalue} and \emph{central emission vertices}, respectively. They have been computed to low orders in perturbation theory in refs.~\cite{Bartels:2009vkz,Lipatov:2010ad,Fadin:2011we,Bartels:2011ge,Dixon:2012yy,Dixon:2014voa,DelDuca:2018hrv}, and they are conjecturally known to all orders from integrability~\cite{Basso:2014pla,DelDuca:2019tur}. The leading-order expressions, which are sufficient to compute $\mathcal{W}_{[4,N-1]}$ to LLA, read: 
\begin{align}
  \chi^{\oplus}(\nu,n) &= \frac{1}{\mi(\nu+\mi0)+\frac{n}{2}}\ab(1+\mathcal{O}(g^2))\:,\nonumber \\
  \omega(\nu,n) &= -2g^2\ab(\psi\ab(1+\mi\nu+\tfrac{|n|}{2})+\psi\ab(1-\mi\nu+\tfrac{|n|}{2})-2\psi(1)-\tfrac{|n|}{2(\nu^2+(n/2)^2)}) \nonumber\\ 
  &\quad+ \mathcal{O}(g^4)\:,\\
  C^{\oplus}(\nu,n,\mu,m)&= \frac{\Gamma\ab(1-\mi\nu-\frac{n}{2})\Gamma\ab(\mi\mu+\frac{m}{2})\Gamma\ab(\mi(\nu-\mu-\mi0)+\frac{m-n}{2})}{\Gamma\ab(1+\mi\nu-\frac{n}{2})\Gamma\ab(-\mi\mu+\frac{m}{2})\Gamma\ab(1-\mi(\nu-\mu)+\frac{m-n}{2})} \ab(1+\mathcal{O}(g^2))\:,\nonumber
\end{align}
and $\chi^{\ominus}(\nu,n) = \chi^{\oplus}(\nu,n)^*$ and $  C^{\ominus}(\nu,n,\mu,m)=  C^{\oplus}(\nu,n,\mu,m)^*$. Here  $\psi(z):=\Gamma'(z)/\Gamma(z)$ is the digamma function. The $\mi0$-prescription removes the divergence of the $\nu$ integral at $n=0$~\cite{DelDuca:2016lad}. 

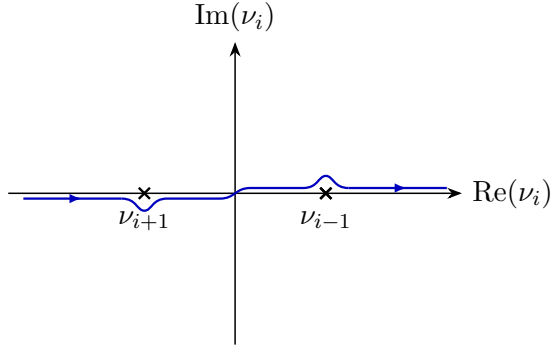
\begin{figure}
  \centering
  \begin{tikzpicture}
    \draw[-{Stealth},line width=0.2mm] (-3,0) -- (3,0) node[right] {$\operatorname{Re}(\nu_{i})$};
    \draw[-{Stealth},line width=0.2mm] (0,-2) -- (0,2) node[above] {$\operatorname{Im}(\nu_{i})$};
    \node[cross=3pt,line width=0.9pt] at(-1.2,0) {};
    \node[cross=3pt,line width=0.9pt] at(1.2,0) {};
    \node[below] at (1.2,-0.1) {$\nu_{i-1}$};
    \node[below] at (-1.2,-0.1) {$\nu_{i+1}$};
       \draw[ 
        decoration={markings, mark=at position 0.6 with {\arrow{Latex[length=1.7mm]}}},
        postaction={decorate},
        line width=0.3mm,
        color=blue!75!black
        ](1.5,0.07) -- (2.8,0.07);  
        \draw[ 
        decoration={markings, mark=at position 0.6 with {\arrow{Latex[length=1.7mm]}}},
        postaction={decorate},
        line width=0.3mm,
        color=blue!75!black
        ](-2.8,-0.07) -- (-1.5,-0.07);
        \node[cross=3pt,line width=0.9pt] at(1.2,0) {};
       \draw[ 
        line width=0.3mm,
        color=blue!75!black
        ]
       (0.2,0.07)-- (0.9,0.07) ..controls (1.1,0.07) and (1.1,0.23).. (1.2,0.23) .. controls (1.3,0.23) and (1.3,0.07).. (1.5,0.07) 
          (-0.2,-0.07)--(-0.9,-0.07) ..controls (-1.1,-0.07) and (-1.1,-0.23).. (-1.2,-0.23) .. controls (-1.3,-0.23) and (-1.3,-0.07).. (-1.5,-0.07) 
        (-0.2,-0.07) .. controls (-0.0,-0.07) and (0.0,0.07) .. (0.2,0.07) ;
\end{tikzpicture}
\caption{The integration contour $\mathcal{C}$ and the poles on the real $\nu_i$ axes in eq.~\eqref{eq:2Reggeon_Ex}, with $\nu_{N-4}=-\pi\tilde{a}$ and $\nu_0=\pi\tilde{a}$ as the boundary cases.} \label{fig:contour}
\end{figure}

It is worth pointing out some nonperturbative features of eq.~\eqref{eq:2Reggeon_Ex}. First, the poles of $\chi^{\oplus}(\nu,0)$ and $\chi^{\ominus}(\nu,0)$ exhibit a nonperturbative shift from $\nu=0$, as suggested by the leading-order result, to $\nu=\pm \pi\tilde{a}$~\cite{Caron-Huot:2013fea}. Together with the $\mi0$-prescription, these shifted poles determine the integration contour $\mathcal{C}$ nonperturbatively, as illustrated in figure~\ref{fig:contour}~\cite{DelDuca:2018hrv,DelDuca:2019tur}. Second, the leading term in the weak-coupling expansion, $\mathcal{W}^{h_4\cdots h_{N-1}}_{[4,N-1]}=(2\pi\mi)^{-1}+\cdots$, is in fact a consequence of the exact nonperturbative residues of $\chi^h$ and $C^h$, together with the exact values of $\omega$ at $\nu=\pi\tilde{a}$ and $n=0$, which are now known as \emph{exact bootstrap conditions}~\cite{Caron-Huot:2013fea,DelDuca:2018hrv}:
\begin{equation}\begin{split}
  \omega(\pm\pi\tilde{a},0)&=0,  \\
  \operatorname{Res}_{\nu=\pm\pi\tilde{a}}\ab(\chi^{\oplus}(\nu,0)\chi^{\ominus}(\nu,0))&=\pm\frac{1}{2\pi g^2}\:, \\
  \operatorname{Res}_{\nu_{1}=\pi\tilde{a}}\ab(\chi^{\oplus}(\nu_{1},0)C^{\oplus}(\nu_{1},0,\nu_{2},n_{2}))&=\mi\chi^{\oplus}(\nu_{2},n_{2}) \:, \\
  \operatorname{Res}_{\nu_{2}=-\pi\tilde{a}}\ab(C^{\oplus}(\nu_{1},n_{1},\nu_{2},0)\chi^{\ominus}(\nu_{2},0))&=-\mi\chi^{\ominus}(\nu_{1},n_{1}) \:, \\
  \operatorname{Res}_{\nu_{2}=\nu_{1}}(C^{\oplus}(\nu_{1},n_{2},\nu_{2},n_{2}))&=-\mi(-1)^{n_2}\me^{\mi\pi\omega(\nu_{2},n_{2})} \:.
\end{split}\end{equation}
Following our earlier convention, we drop the dependence on the helicities when we discuss MHV amplitudes, $ \mathcal{W}_{[4,N-1]}:= \mathcal{W}^{h_4\cdots h_{N-1}}_{[4,N-1]}$.

Other types of Mandelstam regions for scattering amplitudes involving eight or more gluons, however, are not nearly as well understood, with analytic results in MRK only available for the two-loop MHV BDS-normalized amplitudes~\cite{Bargheer:2015djt,DelDuca:2018raq}. The reason for this can be understood by looking at MRK from an effective field theory perspective~\cite{Lipatov:1995pn,Antonov:2004hh}, where MRK amplitudes can be described via the exchange of effective bound states called \emph{Reggeons}. In this description, the Mandelstam regions $[k,l]$ only receive contributions from two-Reggeon exchange. Starting from eight particles, however, $m$-Reggeon exchange with $m\ge 3$ may contribute to other regions~\cite{Lipatov:2009nt,Bartels:2011nz,Caron-Huot:2013fea,Bartels:2020twc}. In the remainder of this paper, we will take first steps toward closing this gap, and we will study $\mathcal{R}_{\varrho,N}$ in the region $\varrho=\{k,l\}$, which receives contributions from both two- and three-Reggeon exchange. We will make heavy use of the effective field theory description of $\mathcal{N}=4$ SYM in MRK developed in refs.~\cite{Bartels:2009vkz,Bartels:2011ge}. Since this approach to MRK has not been prominently used in modern approaches to planar $\mathcal{N}=4$ SYM scattering amplitudes in MRK, we start by reviewing it and discuss how it can be used to explain the known results for two-Reggeon exchange in $\mathcal{N}=4$ SYM.

\section{BFKL effective field theory and two-Reggeon exchange in \texorpdfstring{$\mathcal{N}=4$}{N=4} SYM} \label{sec:twoReggeon}

In this section, we review an effective field theory (EFT) formulation of MRK at LLA, and we will revisit the contribution of two-Reggeon exchange in planar $\mathcal{N}=4$ SYM, mainly following the BFKL approach used in refs.~\cite{Bartels:2009vkz,Bartels:2011ge}. As we shall see, at LLA, we obtain a natural formulation as a two-dimensional EFT.

\subsection{BFKL effective field theory}
\label{sec:Feynman_rules}

The BFKL approach to the multi-Regge limit of amplitudes is not to compute the amplitudes directly, but rather to study their \emph{$s$-channel cuts} and then reconstruct the full amplitudes through dispersion relations. More generally, there is an effective theory description in terms of Reggeons~\cite{Lipatov:1995pn,Antonov:2004hh}. For our purposes, however, only part of this EFT is needed, and it can be formulated in the two-dimensional transverse space. Following ref.~\cite{Caron-Huot:2020vlo}, we shall refer to this reduced theory as the \emph{BFKL effective field theory}, and we review it in this section.

The content of the BFKL-EFT is most easily described in terms of its Feynman rules, which can be summarized as 
follows:
\begin{enumerate}
  \item Since the theory computes the $s$-channel cuts of the amplitudes, the $s$-channel propagators, represented by \emph{horizontal lines}, are cut and put on shell in the original theory. Consequently, they do not contribute as propagators, but rather they provide some $(2\pi\mi)$ factors in the reduced theory. By contrast, the $t$-channel propagators, represented by \emph{vertical lines}, contribute the usual factor in transverse momentum space, as in ordinary QFT:
    \begin{equation}
      \text{propagator:}\quad \begin{tikzpicture}[baseline={([yshift=-0.6ex]current bounding box.center)}]
           \draw[line width=0.3mm] (0,-0.5)--(0,0.5);
           \node[left,scale=0.8] at (0,0) {$\mathbf{x}_{i}$};
           \node[right,scale=0.8] at (0,-0.02) {$\mathbf{x}_{j}$};
    \end{tikzpicture} = \frac{1}{|\mathbf{x}_{ij}|^2}\:. \label{eq: EFT_propagator} 
    \end{equation}
    \item \label{fey_rule2}  There are two types of vertices:
    \begin{align}
    \text{Lipatov vertex:}&\quad \begin{tikzpicture}[baseline={([yshift=-0.6ex]current bounding box.center)}]
           \fill[black] (0,0) circle (0.3ex);
           \draw[line width=0.3mm] (0,0)--(0.5,0) (0,-0.5)--(0,0.5);
           \node[left,scale=0.8] at (0,0) {$\mathbf{x}_{i}$};
           \node[above right,yshift=0.5ex,scale=0.8] at (0,0) {$\mathbf{x}_{a}$};
           \node[below right,yshift=-0.5ex,scale=0.8] at (0,0) {$\mathbf{x}_{b}$};
           \node[right,scale=0.8] at (0.5,0) {$(\oplus/\ominus)$};
    \end{tikzpicture} = 
    \begin{dcases}
            \frac{\mathbf{x}^{\ast}_{ai}\mathbf{x}_{bi}}{\mathbf{x}_{ab}}\,, \quad \text{for positive helicity\,,}\\
            \frac{\mathbf{x}_{ai}\mathbf{x}_{bi}^{\ast}}{\mathbf{x}_{ab}^{\ast}}\,, \quad \text{for negative helicity\,,}
    \end{dcases} \label{eq: EFT_Lipatov_vertex}\\
    \text{peripheral vertex:}&\quad\begin{tikzpicture}[baseline={([yshift=-0.6ex]current bounding box.center)}]
         \fill[black] (0,0) circle (0.3ex);
         \draw[line width=0.3mm] (-0.5,0)node[left,scale=0.8]{$\ominus$}--(0.5,0)node[right,scale=0.8]{$\oplus$} (0,0)--(0,0.5);
         \node[below,yshift=-0.5ex,scale=0.8] at (0,0) {$\mathbf{x}_{i}$};
         \node[above left,yshift=0.5ex,scale=0.8] at (0,0) {$\mathbf{x}_{a}$};
         \node[above right,yshift=0.5ex,scale=0.8] at (0,0) {$\mathbf{x}_{b}$};
    \end{tikzpicture}=\frac{\mathbf{x}_{ai}\mathbf{x}_{bi}^{\ast}}{\mathbf{x}_{ai}^{\ast}\mathbf{x}_{bi}}  
    \quad \text{and} \quad
    \begin{tikzpicture}[baseline={([yshift=-0.6ex]current bounding box.center)}]
         \fill[black] (0,0) circle (0.3ex);
         \draw[line width=0.3mm] (-0.5,0)node[left,scale=0.8]{$\oplus$}--(0.5,0)node[right,scale=0.8]{$\ominus$} (0,0)--(0,0.5);
         \node[below,yshift=-0.5ex,scale=0.8] at (0,0) {$\mathbf{x}_{i}$};
         \node[above left,yshift=0.5ex,scale=0.8] at (0,0) {$\mathbf{x}_{a}$};
         \node[above right,yshift=0.5ex,scale=0.8] at (0,0) {$\mathbf{x}_{b}$};
    \end{tikzpicture}=\frac{\mathbf{x}^{\ast}_{ai}\mathbf{x}_{bi}}{\mathbf{x}_{ai}\mathbf{x}^{\ast}_{bi}} \:,  \label{eq: EFT_peripheral_vertex}
\end{align}
where $\oplus$ and $\ominus$ indicate that the corresponding particles have positive and negative helicity, respectively. Note that the upper peripheral vertex, obtained by flipping the vertex in eq.~\eqref{eq: EFT_peripheral_vertex}, contributes only a factor of unity. Very importantly, when connecting these vertices, \emph{helicity is conserved along each horizontal line} if an outgoing particle of positive helicity is viewed as an incoming particle of negative helicity, and vice versa. We will also encounter a degenerate peripheral vertex where $\mathbf{x}_{a}=\mathbf{x}_{i}$, due to the fact that $\mathbf{p}_{1}=\mathbf{p}_{2}=0$. The Feynman rule for such vertices reads
\begin{equation}
    \begin{tikzpicture}[baseline={([yshift=-0.6ex]current bounding box.center)}]
         \fill[black] (0,0) circle (0.3ex);
         \draw[line width=0.3mm] (-0.5,0)node[left,scale=0.8]{$\ominus$}--(0.5,0)node[right,scale=0.8]{$\oplus$} (0,0)--(0,0.5);
         \node[below,yshift=-0.5ex,scale=0.8] at (0,0) {$\mathbf{x}_{i}$};
         \node[above left,yshift=0.5ex,scale=0.8] at (0,0) {$\mathbf{x}_{i}$};
         \node[above right,yshift=0.5ex,scale=0.8] at (0,0) {$\mathbf{x}_{b}$};
    \end{tikzpicture}=\frac{\mathbf{x}_{bi}^{\ast}}{\mathbf{x}_{bi}} \quad \text{and} \quad 
    \begin{tikzpicture}[baseline={([yshift=-0.6ex]current bounding box.center)}]
         \fill[black] (0,0) circle (0.3ex);
         \draw[line width=0.3mm] (-0.5,0)node[left,scale=0.8]{$\oplus$}--(0.5,0)node[right,scale=0.8]{$\ominus$} (0,0)--(0,0.5);
         \node[below,yshift=-0.5ex,scale=0.8] at (0,0) {$\mathbf{x}_{i}$};
         \node[above left,yshift=0.5ex,scale=0.8] at (0,0) {$\mathbf{x}_{i}$};
         \node[above right,yshift=0.5ex,scale=0.8] at (0,0) {$\mathbf{x}_{b}$};
    \end{tikzpicture}=\frac{\mathbf{x}_{bi}}{\mathbf{x}^{\ast}_{bi}}\:.
\end{equation}
\item There are two ways to add a loop. One can either insert a \emph{bubble} on a vertical line or connect two vertical lines by adding a \emph{BFKL bridge}:
\begin{equation}
 \begin{tikzpicture}[baseline={([yshift=-0.6ex]current bounding box.center)}]
           \fill[black] (0,0) circle (0.3ex);
           \fill[black] (0.6,0) circle (0.3ex);
           \draw[line width=0.3mm] (0,0)--(0.6,0) (0,-0.5)--(0,0.5) (0.6,-0.5)--(0.6,0.5);
           \node[left,scale=0.8] at (0,0) {$\mathbf{x}_{i}$};
           \node[above, yshift=0.5ex,scale=0.8] at (0.3,0) {$\mathbf{x}_{a}$};
           \node[below, yshift=-0.5ex,scale=0.8] at (0.3,0) {$\mathbf{x}_{b}$};
           \node[right,scale=0.8] at (0.6,0) {$\mathbf{x}_{j}$};
    \end{tikzpicture}
    =\frac{\mathbf{x}_{ai}\mathbf{x}^{\ast}_{bi}\mathbf{x}^{\ast}_{aj}\mathbf{x}_{bj}+\text{c.c.}}{|\mathbf{x}_{ab}|^2} \:, \label{eq: EFT_BFKL_bridge}
\end{equation}
where c.c. indicates the addition of the complex conjugate of the foregoing terms.
Note that the BFKL bridge is essentially a sum of products of two Lipatov vertices, with two possible intermediate on-shell gluons.
\item \label{fey_rule4} MRK divides rapidity space into several regions. Depending on the region into which a bubble or BFKL bridge falls, at LLA the coupling constant is modified to
\begin{equation}
\frac{g^2}{2}
\log\frac{p_{i}^{+}p_{j}^{-}}{p_{i}^{-}p_{j}^{+}} \quad \text{for the rapidity region }[\eta_{i},\eta_{j}] \:.
\end{equation}
This is a typical feature of effective field theories: the degrees of freedom that have been integrated out, which are the longitudinal momentum components in our case, are encoded in the effective coupling constant. Furthermore, each $L$-loop insertion in any rapidity region is accompanied by an overall numerical factor of $1/L!$.
\end{enumerate}
Note that we have omitted the color factors associated with the various vertices, since their effects are greatly simplified in the large-$N_c$ limit. Furthermore, we formulate this EFT in two dimensions rather than in $(2-\epsilon)$ dimensions. For one thing, the IR divergences of the original theory are well understood, and our interest lies in the finite quantity $\me^{\mi\pi\tilde{a}\delta^{\varrho}_{N}}\mathcal{R}_{\varrho,N}$. For another, working strictly in two dimensions allows us to use techniques from complex analysis. This has some significant advantages when it comes to evaluating the loop integrals, as we now discuss.

With the Feynman rules introduced above, the Feynman integrals that arise in the EFT take the form
\begin{equation}
   \int_{\mathbb{C}^n} f(z_{1},\ldots,z_{n};c_{1},\ldots,c_{m}) \prod_{i=1}^{n} \dbar^2 z_{i} \:,
\end{equation}
where the measure is normalized as $\dbar^2 z_{i}:= (-1/2\pi\mi)\dif z_{i}\wedge \dif \bar{z}_{i}$, and the integrand $f$ is single-valued with respect to all $z_{i}$ and has poles only coming from terms like $1/(z_i-c_j)$, $1/(\bar z_i-\bar c_j)$, and $1/|z_i-c_j|^2$. Only poles of the last type can give rise to divergences in the integral, as they cannot be compensated by the integral measure $\dbar^2 z_{i}\sim |z_{i}-c_{j}| \dif |z_{i}-c_{j}|\dif \theta_{ij}$ around $z_{i}=c_{j}$. We regulate them by removing a disk of infinitesimal radius $\mu$ centered at $c_{j}$, denoted by $\Delta_{\mu}(c_{j})$, from the complex plane:
\begin{equation}
\int_{\mathbb{C}}  \dbar^2 z_{i} \quad \to \quad \lim_{\mu \to 0} \int_{\mathbb{C} \setminus \Delta_{\mu}(c_{j})}  \dbar^2 z_{i} \: .
\end{equation}
This class of integrals has been studied in detail in the literature, cf., e.g., ref.~\cite{Schnetz:2013hqa}. Viewed as $L$-loop Feynman integrals, they can be evaluated loop by loop and give rise to \emph{single-valued polylogarithms} of weight $L$~\cite{brownSV,BrownSVHPLs,Brown:2013gia,DelDuca:2016lad}. More explicitly, for a single-valued function $f(z)$ with poles at $a_i$ and possibly at infinity, the loop integration gives
\begin{equation}
    \int_{\mathbb{C}} f(z) \dbar^2 z = \Res_{z=\infty} F(z) - \sum_{i} \Res_{z=a_{i}}F(z) \:, \label{eq: svPolylog_int}
\end{equation}
where $F$ is a single-valued, anti-holomorphic primitive of $f$, i.e., $\partial_{\bar z}F=f$. 
For more details, see, e.g., Theorem~2.29 and section~5 of ref.~\cite{Schnetz:2013hqa}, as well as section~4.1 of ref.~\cite{DelDuca:2016lad}. Equation~\eqref{eq: svPolylog_int} is extremely powerful and can be used to evaluate all the integrals appearing in our BFKL-EFT. It was also used in refs.~\cite{DelDuca:2016lad,Marzucca:2018ydt,DelDuca:2018hrv,DelDuca:2019tur} to obtain explicit results through high loop orders starting from the Fourier-Mellin representation in eq.~\eqref{eq:2Reggeon_Ex}.

\subsection{MRK for \texorpdfstring{$\mathcal{N}=4$}{N=4} SYM from the BFKL-EFT} 

We now explain how the BFKL-EFT naturally leads to the Fourier-Mellin representation of the BDS-normalized amplitude from eq.~\eqref{eq:2Reggeon_Ex}.
Recall that the BDS ansatz is essentially the exponentiation of one-loop MHV amplitudes. At four and five points, there are two distinct one-loop Feynman diagrams that can be built from the above Feynman rules:
\begin{align}
    \begin{tikzpicture}[baseline={([yshift=-0.6ex]current bounding box.center)}]
           \draw[line width=0.3mm] (0.25,-0.5)--(0.25,0.5) (-0.25,-0.5)--(-0.25,0.5) ;
           \draw[line width=0.3mm] (-0.6,0.5)--(0.6,0.5) (-0.6,-0.5)--(0.6,-0.5);
           \node[left,xshift=-0.6em,scale=0.8] at (0,0) {$\mathbf{x}_{1}$};
           \node[right,xshift=0.6em,scale=0.8] at (0,-0.02) {$\mathbf{x}_{2}$};
           \node[scale=0.8] at (0,0) {$\mathbf{x}_{a}$};
    \end{tikzpicture} &= \int \frac{|\mathbf{x}_{12}|^2\:\dbar^2 \mathbf{x}_{a}}{|\mathbf{x}_{a1}|^2|\mathbf{x}_{a2}|^2} 
    = 2\log\frac{|\mathbf{x}_{12}|^2}{\mu^2} \label{eq: bub_int} \:, \\
     \begin{tikzpicture}[baseline={([yshift=-0.6ex]current bounding box.center)}]
      \draw[line width=0.3mm] (0.25,-0.5)--(0.25,0.5) (-0.25,-0.5)--(-0.25,0.5) ;
      \draw[line width=0.3mm] (-0.6,0.5)--(0.6,0.5) (-0.6,-0.5)--(0.6,-0.5);
            \fill[black] (0.25,0) circle (0.3ex);
            \draw[line width=0.3mm] (0.25,0)--(0.6,0);
           \node[left,xshift=-0.6em,scale=0.8] at (0,0) {$\mathbf{x}_{1}$};
           \node[above,scale=0.8] at (0.5,0) {$\mathbf{x}_{2}$};
           \node[below,yshift=-0.5ex,scale=0.8] at (0.5,0) {$\mathbf{x}_{3}$};
           \node[scale=0.8] at (0,0) {$\mathbf{x}_{a}$};
    \end{tikzpicture} &= \int \frac{\mathbf{x}_{12}\mathbf{x}_{13}^{\ast}\:\dbar^2 \mathbf{x}_{a}}{|\mathbf{x}_{a1}|^2 \mathbf{x}_{a2}\mathbf{x}_{a3}^{\ast}} =  \log\frac{|\mathbf{x}_{12}|^2|\mathbf{x}_{13}|^2}{|\mathbf{x}_{23}|^2 \mu^2} \:, \label{eq: tri_int}
\end{align}
where we normalized by the inverses of the corresponding tree-level factors. Note that these two integrals are essentially the \emph{bubble} and \emph{triangle} integrals, respectively, since horizontal propagators do not contribute according to the Feynman rules introduced above. Furthermore, up to overall factors, eqs.~\eqref{eq: bub_int} and~\eqref{eq: tri_int} are the gluon Regge trajectory $\omega_{g}(t)$ in eq.~\eqref{eq: gluon_regge_trajectory}, and the one-loop correction to the Lipatov vertex $\omega_{g}(\kappa_{i})$ in eq.~\eqref{eq: one_loop_vertex_correction}.

We can see how gluon reggeization emerges within the framework of the BFKL-EFT. To this end, first note that only the \emph{adjoint} channel of the two-gluon exchange survives in the large-$N_{c}$ limit, which requires that the bubble and the BFKL bridge be organized, order by order in the loop expansion, as (see ref.~\cite{Forshaw:1997dc} for a detailed derivation) 
\begin{align}
     \begin{tikzpicture}[baseline={([yshift=-0.6ex]current bounding box.center)}]
           \draw[line width=0.3mm]  (0.3,-0.7)--(0.3,0.7) (-0.3,-0.7)--(-0.3,0.7);
          \node[rectangle, draw = black, fill = gray!40,font=\footnotesize,inner sep=1.5pt,minimum width=1.1cm,minimum height=0.7cm] at (0,0) {$G_{A}^{(L)}$};
    \end{tikzpicture} \hspace{0.5em} =  \hspace{0.5em}
    \begin{tikzpicture}[baseline={([yshift=-0.6ex]current bounding box.center)}]
           \fill[black] (-0.3,0.55) circle (0.3ex);
           \fill[black] (0.3,0.55) circle (0.3ex);
           \draw[line width=0.3mm]  (0.3,-0.5)--(0.3,0.9) (-0.3,-0.5)--(-0.3,0.9) (-0.3,0.55)--(0.3,0.55);
          \node[rectangle, draw = black, fill = gray!40,font=\footnotesize,inner sep=1.5pt,minimum width=1.1cm] at (0,0) {$G_{A}^{(L-1)}$};
    \end{tikzpicture}   \hspace{0.5em} 
    -   \hspace{0.5em} 
    \begin{tikzpicture}[baseline={([yshift=-0.6ex]current bounding box.center)}]
           \draw[line width=0.3mm]  (0.3,-0.5)--(0.3,0.9) (-0.3,-0.5)--(-0.3,0.9);
            \draw [black,fill=white,line width=0.2mm] (-0.3,0.55) ellipse (0.3em and 0.4em);
          \node[rectangle, draw = black, fill = gray!40,font=\footnotesize,inner sep=1.5pt,minimum width=1.1cm] at (0,0) {$G_{A}^{(L-1)}$};
    \end{tikzpicture}  \hspace{0.5em}
    -   \hspace{0.5em}
   \begin{tikzpicture}[baseline={([yshift=-0.6ex]current bounding box.center)}]
           \draw[line width=0.3mm]  (0.3,-0.5)--(0.3,0.9) (-0.3,-0.5)--(-0.3,0.9);
            \draw [black,fill=white,line width=0.2mm] (0.3,0.55) ellipse (0.3em and 0.4em);
          \node[rectangle, draw = black, fill = gray!40,font=\footnotesize,inner sep=1.5pt,minimum width=1.1cm] at (0,0) {$G_{A}^{(L-1)}$};
    \end{tikzpicture}  \:\:. \label{eq: BFKL_bridge}
\end{align}
We refer to this construction as the \emph{adjoint BFKL ladder}. The adjoint BFKL ladder reduces to a chain of bubbles when attached to a single horizontal propagator, as can be readily verified by direct integration:
\begin{align}
     \begin{tikzpicture}[baseline={([yshift=-0.6ex]current bounding box.center)}]
           \draw[line width=0.3mm]  (0.3,-0.7)--(0.3,0.7) (-0.3,-0.7)--(-0.3,0.7) (0.6,0.7)--(-0.6,0.7);
          \node[rectangle, draw = black, fill = gray!40,font=\footnotesize,inner sep=1.5pt,minimum width=1.1cm,minimum height=0.7cm] at (0,0) {$G_{A}^{(L)}$};
          \node[right,font=\footnotesize] at (0.5,-0.05) {$\mathbf{x}_{j}$};
          \node[left,font=\footnotesize] at (-0.5,-0.05) {$\mathbf{x}_{i}$};
    \end{tikzpicture} \hspace{0.2em} =  \hspace{0.2em}
     -2\log\frac{|\mathbf{x}_{ij}|^2}{\mu^2} \ab(\begin{tikzpicture}[baseline={([yshift=-0.6ex]current bounding box.center)}]
           \draw[line width=0.3mm]  (0.3,-0.7)--(0.3,0.7) (-0.3,-0.7)--(-0.3,0.7) (0.6,0.7)--(-0.6,0.7);
          \node[rectangle, draw = black, fill = gray!40,font=\footnotesize,inner sep=1.5pt,minimum width=1.1cm,minimum height=0.7cm] at (0,0) {$G_{A}^{(L-1)}$};
    \end{tikzpicture} ) \:. \label{eq: BFKL_bridge_recursion}
\end{align}
There is also a flipped version of eq.~\eqref{eq: BFKL_bridge_recursion} where the adjoint BFKL ladder is attached to the bottom end of the diagram. Upon including the coupling constant and the numerical factor specified in rule~\ref{fey_rule4}, the sum over all adjoint BFKL ladders shown above yields the desired factor $s_i^{\omega_g(t_i)}$ in eq.~\eqref{eq:BDSfactorInMRK}, thereby reproducing gluon reggeization: for example,
\begin{align}
      \sum_{L=0}^{\infty} \frac{\eta_{34}^{L}}{L!}   \begin{tikzpicture}[baseline={([yshift=-0.6ex]current bounding box.center)}]
           \draw[line width=0.3mm]  (0.3,-0.7)--(0.3,0.7) (-0.3,-0.7)--(-0.3,0.7) (0.6,0.7)--(-0.6,0.7);
          \node[rectangle, draw = black, fill = gray!40,font=\footnotesize,inner sep=1.5pt,minimum width=1.1cm,minimum height=0.7cm] at (0,0) {$G_{A}^{(L)}$};
          \node[right,font=\footnotesize] at (0.5,-0.05) {$\mathbf{x}_{2}$};
          \node[left,font=\footnotesize] at (-0.5,-0.05) {$\mathbf{x}_{1}$};
    \end{tikzpicture} \hspace{0.2em} &=  \hspace{0.2em}
   \sum_{L=0}^{\infty} \frac{\eta_{34}^{L}}{L!} \ab(-2\log\frac{|\mathbf{x}_{12}|^2}{\mu^2})^{L} \ab(\begin{tikzpicture}[baseline={([yshift=-0.6ex]current bounding box.center)}]
           \draw[line width=0.3mm]  (0.3,-0.7)--(0.3,0.7) (-0.3,-0.7)--(-0.3,0.7) (0.6,0.7)--(-0.6,0.7);
    \end{tikzpicture} ) \sim  
      \ab(\frac{s_{1}}{|\mathbf{p}_{3}\mathbf{p}_{4}|})^{\omega_{g}(t_{1})}
  \:, \label{eq: gluon_reggeization}
\end{align}
where we have used eq.~\eqref{eq:rap_to_s} and neglected some irrelevant factors.

The reduction of the adjoint BFKL ladder to a chain of bubbles is precisely what gives rise to \emph{Regge poles}. Instead, when such a reduction is not possible, \emph{Regge cuts} are required. For instance, in the singlet channel of two-gluon exchange, this reduction fails, because the bubble and the BFKL bridge now enter with coefficients $-1/2$ and $1$, respectively. The resummation of these singlet BFKL ladders is considerably more involved and gives rise to the famous BFKL Pomeron, which corresponds to a Regge cut~\cite{Kuraev:1977fs,Balitsky:1978ic}. On the other hand, this also provides a more explicit explanation of why the four- and five-point amplitudes only exhibit Regge poles in planar $\mathcal{N}=4$ SYM: the adjoint BFKL ladders are inevitably connected to either the top or the bottom horizontal line in both cases.

A crucial change occurs beginning at six points, even for amplitudes where we only encounter adjoint BFKL ladders. The one-loop six-point  amplitude now includes the following Feynman diagram in the BFKL-EFT,
\begin{equation} \label{eq:six_oneloop}
  \begin{tikzpicture}[baseline={([yshift=-0.6ex]current bounding box.center)}]
        \draw[line width=0.3mm] (1,1.0)node[right] {$p_{3}$}--(-1,1.0) (1,-1.0)node[right] {$p_{6}$}--(-1,-1.0);
        \draw[line width=0.3mm] (0.5,1.0)node[circle,draw=black, fill=black, inner sep=0.2ex]{} -- (0.5,-1.0)node[circle,draw=black, fill=black, inner sep=0.2ex]{} 
        (-0.5,1.0)node[circle,draw=black, fill=black, inner sep=0.2ex]{} -- (-0.5,-1.0)node[circle,draw=black, fill=black, inner sep=0.2ex]{};
        \draw[line width=0.3mm] (0.5,0.5)node[circle,draw=black, fill=black, inner sep=0.2ex]{} -- (1.0,0.5)node[right] {$p_{4}$}
        (0.5,-0.5)node[circle,draw=black, fill=black, inner sep=0.2ex]{} -- (1.0,-0.5)node[right] {$p_{5}$};
        \node at (0,0) {$\mathbf{x}_{a}$};
    \end{tikzpicture} =\int\frac{\mathbf{x}_{12}\,\mathbf{x}^{\ast}_{14}\dbar^2 \mathbf{x}_{a}}{|\mathbf{x}_{1a}|^2 \mathbf{x}_{a2}\mathbf{x}_{a4}^{\ast}}
    =\log\frac{|\mathbf{x}_{12}|^{2}|\mathbf{x}_{14}|^{2}}{|\mathbf{x}_{24}|^2\mu^2} \:,
\end{equation}
which yields, up to an overall constant, the phase factor $C^{[4,5]}$ in eq.~\eqref{eq:sixpointBDSphase}. As in eq.~\eqref{eq:sixpointBDSphase}, we can render this integral finite
by subtracting two triangle contributions, which amounts
to making the replacement
\begin{equation}
    \mu^2 \to
    \frac{|\mathbf{x}_{12}\mathbf{x}_{13}^2\mathbf{x}_{14}| }{ |\mathbf{x}_{23}\mathbf{x}_{34}|} \,.
\end{equation}
The resulting expression reproduces the BDS phase $\delta_{6}^{[4,5]}$. However, when the adjoint BFKL ladder is inserted into the rapidity region $[\eta_{4},\eta_{5}]$, it can no longer reduce to a chain of bubble integrals as above, owing to the appearance of the term $\log |\mathbf{x}_{24}|^2$. Note that this Feynman diagram contributes at LLA only in the Mandelstam regions $[4,5]$ and $\{4,6\}$, as shown in ref.~\cite{Lipatov:2009nt}, which agrees with the results obtained from the analytic continuation of the BDS ansatz~\cite{Bartels:2008ce}.

In what follows, we review the construction of Regge cut contributions at LLA within the EFT framework~\cite{Bartels:2009vkz}.
First, we need a procedure for increasing the loop order while preserving finiteness and dual conformal invariance. This can be readily achieved by replacing the adjoint BFKL ladder with the \emph{reduced adjoint BFKL ladder}, defined by
\begin{align}
     \begin{tikzpicture}[baseline={([yshift=-0.6ex]current bounding box.center)}]
           \draw[line width=0.3mm]  (0.3,-0.7)--(0.3,0.7) (-0.3,-0.7)--(-0.3,0.7);
          \node[rectangle, draw = black, fill = gray!40,font=\footnotesize,inner sep=1.5pt,minimum width=1.1cm,minimum height=0.7cm] at (0,0) {$\widehat{G}_{A}^{(L)}$};
          \node[right,font=\footnotesize] at (0.5,-0.05) {$\mathbf{x}_{j}$};
          \node[left,font=\footnotesize] at (-0.5,-0.05) {$\mathbf{x}_{i}$};
          \node[above,font=\footnotesize] at (0,0.3) {$\mathbf{x}_{a}$};
          \node[below,font=\footnotesize] at (0,-0.3) {$\mathbf{x}_{b}$};
    \end{tikzpicture} &=  \hspace{0.5em}
    \begin{tikzpicture}[baseline={([yshift=-0.6ex]current bounding box.center)}]
           \fill[black] (-0.3,0.55) circle (0.3ex);
           \fill[black] (0.3,0.55) circle (0.3ex);
           \draw[line width=0.3mm]  (0.3,-0.5)--(0.3,0.9) (-0.3,-0.5)--(-0.3,0.9) (-0.3,0.55)--(0.3,0.55);
          \node[rectangle, draw = black, fill = gray!40,font=\footnotesize,inner sep=1.5pt,minimum width=1.1cm] at (0,0) {$\widehat{G}_{A}^{(L-1)}$};
    \end{tikzpicture}   \hspace{0.5em} 
    -   \hspace{0.5em} 
    \begin{tikzpicture}[baseline={([yshift=-0.6ex]current bounding box.center)}]
           \draw[line width=0.3mm]  (0.3,-0.5)--(0.3,0.9) (-0.3,-0.5)--(-0.3,0.9);
            \draw [black,fill=white,line width=0.2mm] (-0.3,0.55) ellipse (0.3em and 0.4em);
          \node[rectangle, draw = black, fill = gray!40,font=\footnotesize,inner sep=1.5pt,minimum width=1.1cm] at (0,0) {$\widehat{G}_{A}^{(L-1)}$};
    \end{tikzpicture}  \hspace{0.5em}
    -   \hspace{0.5em}
   \begin{tikzpicture}[baseline={([yshift=-0.6ex]current bounding box.center)}]
           \draw[line width=0.3mm]  (0.3,-0.5)--(0.3,0.9) (-0.3,-0.5)--(-0.3,0.9);
            \draw [black,fill=white,line width=0.2mm] (0.3,0.55) ellipse (0.3em and 0.4em);
          \node[rectangle, draw = black, fill = gray!40,font=\footnotesize,inner sep=1.5pt,minimum width=1.1cm] at (0,0) {$\widehat{G}_{A}^{(L-1)}$};
    \end{tikzpicture} \hspace{0.5em} + 2\log\frac{|\mathbf{x}_{ij}|^2}{\mu^2}
    \ab(\begin{tikzpicture}[baseline={([yshift=-0.6ex]current bounding box.center)}]
           \draw[line width=0.3mm]  (0.3,-0.5)--(0.3,0.9) (-0.3,-0.5)--(-0.3,0.9);
          \node[rectangle, draw = black, fill = gray!40,font=\footnotesize,inner sep=1.5pt,minimum width=1.1cm] at (0,0) {$\widehat{G}_{A}^{(L-1)}$};
    \end{tikzpicture}) \nonumber \\
    &=  \hspace{0.5em}
    \begin{tikzpicture}[baseline={([yshift=-0.6ex]current bounding box.center)}]
           \fill[black] (-0.3,0.55) circle (0.3ex);
           \fill[black] (0.3,0.55) circle (0.3ex);
           \draw[line width=0.3mm]  (0.3,-0.5)--(0.3,0.9) (-0.3,-0.5)--(-0.3,0.9) (-0.3,0.55)--(0.3,0.55);
          \node[rectangle, draw = black, fill = gray!40,font=\footnotesize,inner sep=1.5pt,minimum width=1.1cm] at (0,0) {$\widehat{G}_{A}^{(L-1)}$};
    \end{tikzpicture}   \hspace{0.5em} - 2\log\frac{|\mathbf{x}_{ai}|^2|\mathbf{x}_{aj}|^2}{|\mathbf{x}_{ij}|^2\mu^2} \ab(
      \begin{tikzpicture}[baseline={([yshift=-0.6ex]current bounding box.center)}]
           \draw[line width=0.3mm]  (0.3,-0.5)--(0.3,0.9) (-0.3,-0.5)--(-0.3,0.9);
          \node[rectangle, draw = black, fill = gray!40,font=\footnotesize,inner sep=1.5pt,minimum width=1.1cm] at (0,0) {$\widehat{G}_{A}^{(L-1)}$};
    \end{tikzpicture}
    ) \:,
        \label{eq: reduce_BFKL_bridge}
\end{align}
which amounts simply to subtracting from the original ladder the Regge-pole contribution as if it were attached to a single horizontal line (cf. eq.~\eqref{eq: BFKL_bridge_recursion}). Second, we replace the coupling constant $g^2\eta_{i+3,i+4}$ by its dual-conformal completion:
\begin{equation} \label{eq:reg_prop_sign}
  g^2\eta_{i+3,i+4}\to \begin{cases}
    -g^2(\log\tau_i+\mi\pi)\,, & \text{if }  \varrho_{i+3}\varrho_{i+4}>0  \:, \\
    -g^2\log\tau_i\,, & \text{if } \varrho_{i+3}\varrho_{i+4}<0  \:.
  \end{cases} 
\end{equation}
With these two ingredients in place, we can now define the \emph{reduced adjoint BFKL Green's function} by
\begin{equation}\label{eq:reduced_BFKL}
     \begin{tikzpicture}[baseline={([yshift=-0.6ex]current bounding box.center)}]
           \draw[line width=0.3mm]  (0.3,-0.7)--(0.3,0.7) (-0.3,-0.7)--(-0.3,0.7);
          \node[rectangle, draw = black, fill = gray!40,font=\footnotesize,inner sep=1.5pt,minimum width=1.1cm,minimum height=0.7cm] at (0,0) {$\widehat{G}_{A}$};
          \node[right,font=\footnotesize] at (0.5,-0.05) {$\mathbf{x}_{j}$};
          \node[left,font=\footnotesize] at (-0.5,-0.05) {$\mathbf{x}_{i}$};
          \node[above,font=\footnotesize] at (0,0.3) {$\mathbf{x}_{a}$};
          \node[below,font=\footnotesize] at (0,-0.3) {$\mathbf{x}_{b}$};
    \end{tikzpicture} = \sum_{L=0}^{\infty} \frac{\tilde{\eta}^{L}}{L!} 
    \ab( \begin{tikzpicture}[baseline={([yshift=-0.6ex]current bounding box.center)}]
           \draw[line width=0.3mm]  (0.3,-0.7)--(0.3,0.7) (-0.3,-0.7)--(-0.3,0.7);
          \node[rectangle, draw = black, fill = gray!40,font=\footnotesize,inner sep=1.5pt,minimum width=1.1cm,minimum height=0.7cm] at (0,0) {$\widehat{G}_{A}^{(L)}$};
    \end{tikzpicture} ) \:,
\end{equation}
where the dual-conformal completions of the factors $g^2\eta_{i+3,i+4}$ are collectively denoted by $\tilde{\eta}$. It is also evident that this Green's function satisfies the well-known reduced \emph{BFKL equation},
\begin{equation}
  \partial_{\tilde{\eta}} \widehat{G}_{A} = \mathcal{K}_{A} \mathbin{\star} \widehat{G}_{A} \:,
\end{equation}
where $\mathcal{K}_{A}$ denotes the reduced BFKL kernel, an integral operator whose action increases the loop order by one: $\mathcal{K}_{A} \mathbin{\star} \widehat{G}^{(L)}_{A} = \widehat{G}^{(L+1)}_{A}$. By solving the BFKL equation, one can resum the large logarithms that arise from MRK.

For the simplest case involving a (conformal) Regge cut, the quantity $\mathcal{R}_{[4,5],6}\me^{\pi\mi\tilde{a}\delta_{6}^{[4,5]}}$ is then given at LLA by
\begin{equation}
\mathcal{R}_{[4,5],6}\me^{\pi\mi\tilde{a}\delta_{6}^{[4,5]}}= 2\pi \mi  \underbrace{\ab(\,
  \begin{tikzpicture}[baseline={([yshift=-0.6ex]current bounding box.center)}]
        \draw[line width=0.3mm] (1,1.0)--(-1,1.0) (1,-1.0)--(-1,-1.0);
        \draw[line width=0.3mm] (0.5,1.0)node[circle,draw=black, fill=black, inner sep=0.2ex]{} -- (0.5,-1.0)node[circle,draw=black, fill=black, inner sep=0.2ex]{} 
        (-0.5,1.0)node[circle,draw=black, fill=black, inner sep=0.2ex]{} -- (-0.5,-1.0)node[circle,draw=black, fill=black, inner sep=0.2ex]{};
        \draw[line width=0.3mm] (0.5,0.5)node[circle,draw=black, fill=black, inner sep=0.2ex]{} -- (1.0,0.5)
        (0.5,-0.5)node[circle,draw=black, fill=black, inner sep=0.2ex]{} -- (1.0,-0.5);
           \node[rectangle, draw = black, fill = gray!40,font=\footnotesize,inner sep=1.5pt,minimum width=1.3cm,minimum height=0.6cm] at (0,0) {$\widehat{G}_{A}$};
    \end{tikzpicture}\,+\cdots) }_{=:\mathcal{W}_{[4,5]}}\:, \label{eq:sixpointMRK_diag}
\end{equation}
where the factor $2\pi\mi$ signifies the $s$-channel cut of the associated Feynman diagram and the dots indicate higher-order terms. As argued in ref.~\cite{Caron-Huot:2013fea}, the higher-order contributions can always be absorbed into corrections to the BFKL kernel and the various vertices in the Feynman diagram above. Schematically, this can be written as
\begin{align} 
  \mathcal{W}_{[4,5]} &= \langle\chi_{f}|\me^{-(\log\tau+\mi\pi)\widetilde{\mathcal{K}}_{A}}|\chi_{i}\rangle  \nonumber \\
   &=g^2 \sum_{n=-\infty}^{\infty}\ab(\frac{z}{\bar{z}})^{n/2}\int_{-\infty}^{\infty}  \frac{\dif \nu}{2\pi} |z|^{2\mi\nu}
    \chi^{\oplus}(\nu,n)\me^{-\omega(\nu,n)(\log\tau+\mi\pi)}\chi^{\ominus}(\nu,n) \:,  \label{eq:sixpoint_MRK}
\end{align}
where $\chi_i$ and $\chi_f$, commonly referred to as the \emph{impact factors}, represent the upper and lower parts of the Feynman diagram including all-order corrections, respectively, while $\widetilde{\mathcal{K}}_A$ denotes the all-order reduced adjoint BFKL kernel. We also suppress the subscripts on $z_i$ and $\log\tau_i$, since there is only one of each in the present case. The second equality is an inverse Fourier-Mellin transform that simply expresses this matrix element in the eigenbasis of the operator $\widetilde{\mathcal{K}}_A$, whose eigenfunctions are labeled by the conformal quantum numbers $(\nu,n)\in\mathbb{R}\times\mathbb{Z}$. This is exactly how the Fourier-Mellin representation \eqref{eq:2Reggeon_Ex} was originally obtained at leading order (for $N=6$). In particular, if we work at LLA, then at every loop order we can compute the contribution from the Fourier-Mellin integral in eq.~\eqref{eq:2Reggeon_Ex} via Feynman diagrams in the EFT. This result is of course well known, but it serves as a motivation for our method to compute the contribution from three-Reggeon exchange in the next section.

\section{Three-Reggeon exchange in the zigzag octagon}
\label{sec:three-reggeon}

In the previous section, we saw that adjoint two-Reggeon exchange, which is generally associated with gluon reggeization and Regge poles, gives rise to a Regge cut in planar $\mathcal{N}=4$ SYM starting at six points. 
In this section, we present our main results, namely, we discuss how to compute the contributions at LLA in Mandelstam regions where both two- and three-Reggeon are nonzero. We start by some general considerations concerning multi-Reggeon exchange from the EFT perspective, and we specialize to three-Reggeon exchange in subsequent subsections.

\subsection{Multi-Reggeon exchange to LLA}
\begin{figure}
  \centering
\begin{tikzpicture}[baseline={([yshift=-0.6ex]current bounding box.center)}]
\draw[line width=0.3mm] (-3,2.5) -- (1.5,2.5);
\draw[line width=0.3mm] (-3,-4) -- (1.5,-4);
\draw[line width=0.3mm] (-1.5,2.5)node[circle,draw=black, fill=black, inner sep=0.2ex] {} -- (-1.5,-4)node[circle,draw=black, fill=black, inner sep=0.2ex] {};
\draw[line width=0.3mm] (-0.5,2.5)node[circle,draw=black, fill=black, inner sep=0.2ex] {} -- (-0.5,-4)node[circle,draw=black, fill=black, inner sep=0.2ex] {};
\draw[line width=0.3mm] (-0.5,1.5) -- (2,1.5);
\draw[line width=0.3mm] (-0.5,-3) -- (2,-3);
\draw[line width=0.3mm] (0.5,1.5)node[circle,draw=black, fill=black, inner sep=0.2ex] {} -- (0.5,-3)node[circle,draw=black, fill=black, inner sep=0.2ex] {};
\draw[line width=0.3mm] (0.5,0.5) -- (2.5,0.5);
\draw[line width=0.3mm] (0.5,-2) -- (2.5,-2);
\draw[line width=0.3mm] (1.5,0.5)node[circle,draw=black, fill=black, inner sep=0.2ex] {} -- (1.5,-2)node[circle,draw=black, fill=black, inner sep=0.2ex] {};
\draw[line width=0.3mm] (1.5,-0.0)node[circle,draw=black, fill=black, inner sep=0.2ex] {} -- (2.5,-0.0);
\draw[line width=0.3mm] (1.5,-1.5)  node[circle,draw=black, fill=black, inner sep=0.2ex] {}-- (2.5,-1.5);
\node at (2.9,-0.15) {$\ddots$};
\node at (2.9,-1.15) {$\iddots$};
\node at (-3.5,2.5) {$2$};
\node[right] at (1.5,2.5) {$3\quad (+)$};
\node[right] at (2.0,1.5) {$4\quad (-)$};
\node at (-3.5,-4) {$1$};
      \draw[
    black,
    fill=gray!150!black,
    rounded corners=4pt,
    line width=0.3mm
  ] (-1.7,-1.0) rectangle (3.5,-0.55);
\node[right] at (1.5,-4) {$N\quad (+)$};
\node[right] at (2.0,-3) {$N{-}1\quad (-)$};
\node[right] at (2.5,0.5) {$5\quad (+)$};
\node[right] at (2.5,-2) {$N{-}2\quad (-)$};
\node[circle,draw=black, fill=black, inner sep=0.2ex] at (-0.5,1.5) {};
\node[circle,draw=black, fill=black, inner sep=0.2ex] at (0.5,-2) {};
\node[circle,draw=black, fill=black, inner sep=0.2ex] at (-0.5,-3) {};
\node[circle,draw=black, fill=black, inner sep=0.2ex] at (0.5,0.5) {};
\node at (-2.5,-1) {$\mathbf{x}_1$};
\node at (0.5,2) {$\mathbf{x}_2$};
\node at (1.5,1) {$\mathbf{x}_{3}$};
\node at (1.5,-2.5) {$\mathbf{x}_{N-3}$};
\node at (0.5,-3.5) {$\mathbf{x}_{N-2}$};
\node at (-0.5,3){$\mathbf{x}_{1}$};
\node at (-0.5,-4.5){$\mathbf{x}_{1}$};
\end{tikzpicture} 
\caption{ \label{fig:multi-Reggeon_exchange}Schematic representation of a Feynman diagram in the BFKL-EFT describing the LLA contribution of $m$-Reggeon-exchange. The signs in parentheses are the energy signs $\varrho_i$. The gray box represents the insertion of a reduced BFKL Green's function. }
\end{figure}
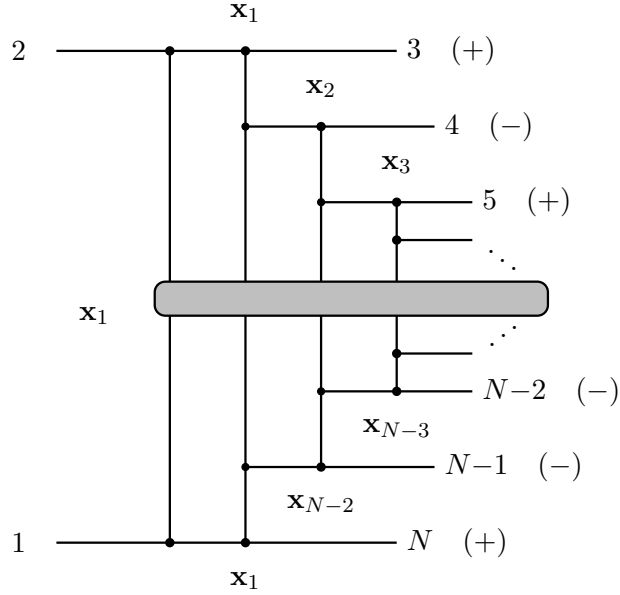

The LLA of multi-Reggeon exchange in planar $\mathcal{N}=4$ SYM can be described very easily from the EFT perspective. In appendix~A of ref.~\cite{Lipatov:2009nt} it was argued that $m$-Reggeon exchange can be obtained by generalizing the diagram from six points by successively grafting smaller boxes onto the right-hand side of the preceding box, as illustrated in figure~\ref{fig:multi-Reggeon_exchange}, where the $m$-Reggeon exchange is generated by adding loops within the central gray-box region according to rule~\ref{fey_rule4}. To access the corresponding $s$-channel cuts at LLA, the energy signs $(\varrho_i)_{3\leq i\leq N}$ must follow the alternating pattern\footnote{In ref.~\cite{Lipatov:2009nt}, what is actually shown is that the energy signs in each half have to follow the alternating pattern \emph{individually} to access the multi-Reggeon exchange. In other words, the multi-Reggeon exchange would exist in two inequivalent regions: $({+}{-}{+}{-}\cdots{-}{+}{-}{+})$ and $({+}{-}{+}{-}\cdots{+}{-}{+}{-})$, and we will leave a detailed study of the second kind of regions to elsewhere.} $({+}{-}{+}{-}\cdots{-}{+}{-}{+})$, as also indicated in the diagram. Following ref.~\cite{Caron-Huot:2013fea}, we shall refer to such kinematic regions as \emph{zigzag regions} and to the corresponding $N$-point amplitudes as \emph{zigzag $N$-gons}.

It is evident that the $m$-Reggeon exchange can occur only for multiplicities $N\geq 2(m+1)$. Therefore, the first amplitude exhibiting a Regge cut due to three-Reggeon exchange is the zigzag octagon $\mathcal{R}_{\{4,7\},8}\me^{\pi\mi\tilde{a}\delta_{8}^{\{4,7\}}}$. Since the exchange of $m$ Reggeons requires at least a $(m-1)$-fold $s$-channel cut, it carries a factor of $(2\pi\mi)^{m-1}$. 
We can now immediately use the Feynman rules of the BFKL-EFT in section~\ref{sec:Feynman_rules} to write a Feynman integral representation in a two-dimensional field theory that evaluates the LLA of the diagram in figure~\ref{fig:multi-Reggeon_exchange} (with an appropriate number of insertions from the reduced BFKL ladder represented by the gray box).

At this point, we have to address an important point. The BFKL-EFT allows us to access the LLA of the contribution from $m$-Reggeon exchange. From the Feynman rules in section~\ref{sec:Feynman_rules}, it follows that the LLA of an $L$-loop diagram for $m$-Reggeon exchange like in figure~\ref{fig:multi-Reggeon_exchange} takes the schematic form
\begin{equation}
(2\pi \mi)^{m-1}\,\sum_{(r_1,\ldots,r_{N-5})}X_{\varrho}^{r_1,\ldots, r_{N-5}}(z_1,\ldots,z_{N-5})\,\prod_{i=1}^{N-5}\log^{r_i}\tau_i\,
\end{equation}
with the constraint $r_1+\ldots+r_{N-5}=L-m+1$. The dependence of $X_\varrho^{r_1,\ldots,r_{N-5}}$ on the transverse coordinates is determined entirely by the Mandelstam region $\varrho$. From this equation we conclude that the LLA of $m$-Reggeon exchange contributes to the N$^{m-2}$LLA of the BDS-normalized amplitude $\mathcal{R}_{\varrho,N}\me^{\pi\mi\tilde{a}\delta_{N}^{\varrho}}$. Hence, in order to turn the computation of the LLA of $m$-Reggeon exchange into a prediction for the BDS-normalized amplitude, we need to combine it with the contributions from $p$-Reggeon exchange with $p<m$ and contributing to the same order in the logarithmic expansion. 
Since in our paper we are concerned with three-Reggeon exchange at LLA in the region $\{k,l\}$, we also need to include contributions from two-Reggeon exchange at NLLA. So far NLLA  results have only been computed for the Mandelstam regions $[k,l]$~\cite{Fadin:2011we,Dixon:2011pw,Dixon:2012yy,DelDuca:2018hrv,Marzucca:2018ydt,DelDuca:2018raq,DelDuca:2019tur}. We therefore first need to discuss how to obtain NLLA results for the two-Reggeon exchange in the regions $\{k,l\}$.

\subsection{Two-Reggeon exchange in the zigzag octagon}

In this section, we explain how to obtain results for the zigzag octagon $\mathcal{R}_{\{4,7\},8}
    \me^{\pi\mi\tilde{a}\delta_{8}^{\{4,7\}}}$, which receives contributions from both two- and three-Reggeon exchange.
Since the three-Reggeon exchange contribution always comes with a factor of $(2\pi \mi)^2$, we write
\begin{equation} \label{eq:dec_oct}
    \mathcal{R}_{\{4,7\},8}
    \me^{\pi\mi\tilde{a}\delta_{8}^{\{4,7\}}}
    =
    1+(2\pi\mi)\mathcal{W}_{\{4,7\},8}
    +(2\pi\mi)^2\mathcal{T}_{\{4,7\},8} \,,
\end{equation}
where $\mathcal{W}_{\varrho,N}$ and $\mathcal{T}_{\varrho,N}$ denote the contributions for the $N$-point BDS-normalized amplitude in the Mandelstam region $\varrho$ from two- and three-Reggeon exchanges, respectively, normalized by an appropriate power of $2\pi \mi$. If clear from the context, we will often drop the subscript $N$, and simply write $\mathcal{W}_{\varrho}$ and $\mathcal{T}_{\varrho}$ instead of $\mathcal{W}_{\varrho,N}$ and $\mathcal{T}_{\varrho,N}$. The contribution $\mathcal{T}_{\{4,7\},8}$ at LLA can be computed using the Feynman rules for the BFKL-EFT from section~\ref{sec:Feynman_rules}. However, so far the contributions from $\mathcal{W}_{\{4,7\},8}$ are only available at two-loop order beyond LLA. In the following we present a proposal for an all-order expression of $\mathcal{W}_{\{4,7\},8}$ as a linear combination of $\mathcal{W}_{[k,l],8}$, which are known to all orders from eq.~\eqref{eq:2Reggeon_Ex}.

We begin by reviewing some basic properties of the functions $\mathcal{R}_{\varrho,N}^{(L)}$, obtained by first analytically continuing the BDS-normalized amplitude $R_{N}^{(L)}$ to the
Mandelstam region $\varrho$ and then taking its multi-Regge limit. The resulting function admits a natural grading by powers of $2\pi\mi$ and can be written as
\begin{equation}
    \mathcal{R}_{\varrho,N}^{(L)}
    =
    \sum_{k=1}^{2L}(2\pi\mi)^k X_{\varrho,N}^{(L),k} \,,
\end{equation}
where each coefficient $X_{\varrho,N}^{(L),k}$ is a single-valued
polylogarithmic function of weight $2L-k$.

As a consequence of the first-entry condition~\cite{Gaiotto:2011dt}, the functions
$X_{\varrho,N}^{(L),1}$ associated with different Mandelstam regions
$\varrho$ are related, as shown in ref.~\cite{Bargheer:2015djt}. The relation
relevant for our purposes is
\begin{equation} \label{eq:Xrel1}
    X_{\{k,l\},N}^{(L),1}=X_{[k,l],N}^{(L),1}-X_{[k+1,l],N}^{(L),1}-X_{[k,l-1],N}^{(L),1}+X_{[k+1,l-1],N}^{(L),1}\:.
\end{equation} 
Furthermore, it follows from the decomposition in eq.~\eqref{eq:dec_oct} that $X_{\varrho,8}^{(L),1}$ contributes exclusively to the two-Reggeon exchange term $\mathcal{W}_{\varrho,8}$ and determines its constant term in an expansion in $2\pi\mi$. Combining this observation with eq.~\eqref{eq:Xrel1}, we obtain the constraint 
\begin{equation} \label{eq:W_constraint}
  \mathcal{W}_{\{4,7\},8}=\mathcal{W}_{[4,7],8}-\mathcal{W}_{[4,6],8}-\mathcal{W}_{[5,7],8}+\mathcal{W}_{[5,6],8}+\ldots\:,
\end{equation}
where the dots indicate terms proportional to additional powers of $2\pi\mi$. At one loop, this yields the BDS-phase relation in eq.~\eqref{eq: BDSphase_relation}, since $\mathcal{W}_{\varrho,N}^{(1)}=\delta_N^\varrho/2$.

We can constrain $\mathcal{W}_{\{4,7\},8}$ further by studying the soft behavior of both the
BDS-normalized amplitudes and the BDS phases. The amplitudes satisfy
\begin{equation} \label{eq:R_soft}
  \begin{aligned}
    &\mathcal{R}_{\{4,7\},8} \to 1,\:
    \mathcal{R}_{[4,7],8} \to \mathcal{R}_{[5,7],8},\:
    \mathcal{R}_{[4,6],8} \to \mathcal{R}_{[5,6],8},\:
    && \text{as $\mathbf{p}_{4}\to0$, or equivalently $z_{1}\to0$}\:,
    \\
    &\mathcal{R}_{\{4,7\},8} \to 1, \:
    \mathcal{R}_{[4,7],8} \to \mathcal{R}_{[4,6],8}, \:
    \mathcal{R}_{[5,7],8} \to \mathcal{R}_{[5,6],8}, \:
  && \text{as $\mathbf{p}_{6}\to0$, or equivalently $z_{3}\to\infty$}\:.
\end{aligned}
\end{equation}
The corresponding BDS phases obey
\begin{equation} \label{eq:delta_soft}
  \begin{aligned}
    \lim_{z_{1}\to0}\delta_{8}^{[4,7]} &= \log|z_{1}|^2+\delta_{8}^{[5,7]}\:,
    &
    \lim_{z_{1}\to0}\delta_{8}^{[4,6]} &= \log|z_{1}|^2+\delta_{8}^{[5,6]}\:,
    \\
    \lim_{z_{3}\to\infty}\delta_{8}^{[4,7]} &= -\log|z_{3}|^2+\delta_{8}^{[4,6]}\:,
    &
  \lim_{z_{3}\to\infty}\delta_{8}^{[5,7]} &= -\log|z_{3}|^2+\delta_{8}^{[5,6]}\:.
\end{aligned}
\end{equation}

Equations~\eqref{eq:W_constraint},~\eqref{eq:R_soft} and~\eqref{eq:delta_soft} put strong constraints on the possible functional forms for $\mathcal{W}_{\{4,7\},8}$. 
In order to satisfy all these constraints, we propose
\begin{equation}
    \label{eq:two_reggeon_sub}
    \mathcal{W}_{\{4,7\},8}
    =
    \ab[
        \mathcal{W}_{[4,7],8}
        -\mathcal{W}_{[4,6],8}|z_{3}|^{-2\pi\mi\tilde{a}}
        -\mathcal{W}_{[5,7],8}|z_{1}|^{2\pi\mi\tilde{a}}
        +\mathcal{W}_{[5,6],8}
         \ab|\frac{z_{1}}{z_{3}}|^{2\pi\mi\tilde{a}}
    ]
    \Bigg|_{\log\tau_{1,3}\to\log\tau_{1,3}-\mi\pi}
    \:.
\end{equation}
Here, the prescription in eq.~\eqref{eq:reg_prop_sign} has also been
incorporated through the replacements
$\log\tau_i\to\log\tau_i-\mi\pi$ for $i=1,3$. We expect eq.~\eqref{eq:two_reggeon_sub} to be valid to all orders in perturbation theory. In particular, the $\mathcal{W}_{[k,l],8}$ appearing in the right-hand side of eq.~\eqref{eq:two_reggeon_sub} are given as a Fourier-Mellin integral in eq.~\eqref{eq:2Reggeon_Ex}, and all the ingredients entering the Fourier-Mellin integral are (conjecturally) known to all orders from integrability~\cite{Basso:2014pla,DelDuca:2019tur}. We can then use the techniques developed in ref.~\cite{DelDuca:2016lad} to evaluate $\mathcal{W}_{\{4,7\},8}$ to any desired order in perturbation theory in terms of single-valued polylogarithms, including contributions beyond LLA. In appendix \ref{app:B}, we review how to connect the perturbative expansion of $\mathcal{W}_{\{4,7\},8}$ to available results in the literature, in particular from refs.~\cite{Marzucca:2018ydt,DelDuca:2019tur}.

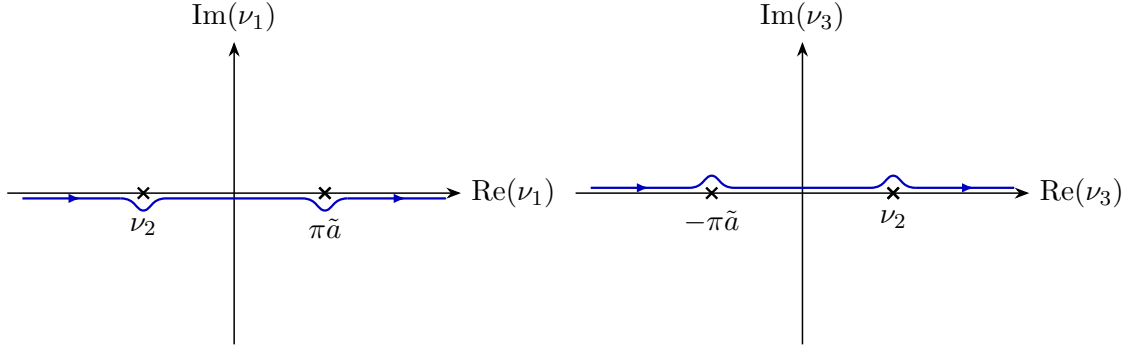
\begin{figure}
  \centering
  \begin{tikzpicture}
    \draw[-{Stealth},line width=0.2mm] (-3,0) -- (3,0) node[right] {$\operatorname{Re}(\nu_{1})$};
    \draw[-{Stealth},line width=0.2mm] (0,-2) -- (0,2) node[above] {$\operatorname{Im}(\nu_{1})$};
    \node[cross=3pt,line width=0.9pt] at(-1.2,0) {};
    \node[cross=3pt,line width=0.9pt] at(1.2,0) {};
    \node[below] at (1.2,-0.2) {$\pi\tilde{a}$};
    \node[below] at (-1.2,-0.2) {$\nu_{2}$};
       \draw[ 
        decoration={markings, mark=at position 0.6 with {\arrow{Latex[length=1.7mm]}}},
        postaction={decorate},
        line width=0.3mm,
        color=blue!75!black
        ](1.5,-0.07) -- (2.8,-0.07);  
        \draw[ 
        decoration={markings, mark=at position 0.6 with {\arrow{Latex[length=1.7mm]}}},
        postaction={decorate},
        line width=0.3mm,
        color=blue!75!black
        ](-2.8,-0.07) -- (-1.5,-0.07);
        \node[cross=3pt,line width=0.9pt] at(1.2,0) {};
       \draw[ 
        line width=0.3mm,
        color=blue!75!black
        ]
       (0.2,-0.07)-- (0.9,-0.07) ..controls (1.1,-0.07) and (1.1,-0.23).. (1.2,-0.23) .. controls (1.3,-0.23) and (1.3,-0.07).. (1.5,-0.07) 
          (-0.2,-0.07)--(-0.9,-0.07) ..controls (-1.1,-0.07) and (-1.1,-0.23).. (-1.2,-0.23) .. controls (-1.3,-0.23) and (-1.3,-0.07).. (-1.5,-0.07) 
        (-0.2,-0.07) -- (0.2,-0.07) ;
\end{tikzpicture}
\begin{tikzpicture}
    \draw[-{Stealth},line width=0.2mm] (-3,0) -- (3,0) node[right] {$\operatorname{Re}(\nu_{3})$};
    \draw[-{Stealth},line width=0.2mm] (0,-2) -- (0,2) node[above] {$\operatorname{Im}(\nu_{3})$};
    \node[cross=3pt,line width=0.9pt] at(-1.2,0) {};
    \node[cross=3pt,line width=0.9pt] at(1.2,0) {};
    \node[below] at (1.2,-0.1) {$\nu_{2}$};
    \node[below] at (-1.2,-0.1) {$-\pi\tilde{a}$};
       \draw[ 
        decoration={markings, mark=at position 0.6 with {\arrow{Latex[length=1.7mm]}}},
        postaction={decorate},
        line width=0.3mm,
        color=blue!75!black
        ](1.5,0.07) -- (2.8,0.07);  
        \draw[ 
        decoration={markings, mark=at position 0.6 with {\arrow{Latex[length=1.7mm]}}},
        postaction={decorate},
        line width=0.3mm,
        color=blue!75!black
        ](-2.8,0.07) -- (-1.5,0.07);
        \node[cross=3pt,line width=0.9pt] at(1.2,0) {};
       \draw[ 
        line width=0.3mm,
        color=blue!75!black
        ]
       (0.2,0.07)-- (0.9,0.07) ..controls (1.1,0.07) and (1.1,0.23).. (1.2,0.23) .. controls (1.3,0.23) and (1.3,0.07).. (1.5,0.07) 
          (-0.2,0.07)--(-0.9,0.07) ..controls (-1.1,0.07) and (-1.1,0.23).. (-1.2,0.23) .. controls (-1.3,0.23) and (-1.3,0.07).. (-1.5,0.07) 
        (-0.2,0.07) -- (0.2,0.07) ;
\end{tikzpicture}
\caption{The deformed integration contours $\mathcal{C}'$ for the integrations over $\nu_1$ (left) and $\nu_3$ (right) in eq.~\eqref{eq:2Reggeon_47}. The contour for the integration over $\nu_2$ is the same as that shown in figure~\ref{fig:contour}.} \label{fig:deformed_contour}
\end{figure}

We note that the expression for $\mathcal{W}_{\{4,7\},8}$ proposed in eq.~\eqref{eq:two_reggeon_sub} admits a compact Fourier-Mellin integral representation: 
\begin{equation}
  \mathcal{W}_{\{4,7\},8}= g^2 \prod_{r=1}^{3}\ab[\sum_{n_r}\ab(\frac{z_{r}}{\bar{z}_{r}})^{n_{r}/2}\int_{\mathcal{C}'}\frac{\dif \nu_{r}}{2\pi}]  \frac{  \chi^{\oplus}_{1}\,C^{\oplus}_{12} \, C^{\oplus}_{23}\,\chi^{\ominus}_{3}}{\tau_{1}^{\omega_1}(-\tau_{2}+\mi 0)^{\omega_2}\tau_{3}^{\omega_3}}
 \:, \label{eq:2Reggeon_47} 
\end{equation}
where the integration contour $\mathcal{C}'$ is defined as shown in figure~\ref{fig:deformed_contour}. 
The additional factors $|z_{3}|^{-2\pi\mi\tilde{a}}$, $|z_{1}|^{2\pi\mi\tilde{a}}$, and $|z_{1}/z_{3}|^{2\pi\mi\tilde{a}}$ in eq.~\eqref{eq:two_reggeon_sub}, introduced to ensure the correct soft behavior, can be understood as the consequence of contour deformations, in analogy with the cases discussed in refs.~\cite{DelDuca:2018hrv}. We see that our proposal takes a form very similar to the two-Reggeon contribution in the regions $[k,l]$ given in eq.~\eqref{eq:2Reggeon_Ex}, the only difference lying in the phase factors carried by the $\tau_i$ and in the form of the integration contour. The compact representation in eq.~\eqref{eq:2Reggeon_47} also suggests an immediate extension to non-MHV helicity configurations, by simply using the appropriate impact factors and central emission vertices.

Together with our recipe for computing $\mathcal{T}_{\{4,7\},8}$ from the BFKL-EFT to LLA, this provides all the ingredients needed to obtain analytic results for $\mathcal{R}_{\{4,7\},8}\me^{\pi\mi\tilde{a}\delta_{8}^{\{4,7\}}}$ up to NLLA to any desired order in perturbation theory. However, these predictions crucially rely on our conjectured form for $\mathcal{W}_{\{4,7\},8}$ in eqs.~\eqref{eq:two_reggeon_sub} and~\eqref{eq:2Reggeon_47}. At one loop, this prescription fixes our choice of the BDS phase given in eq.~\eqref{eq: BDSphase_relation}. In the remainder of this section we show that our conjecture for $\mathcal{W}_{\{4,7\},8}$ is consistent with the known perturbative results for the eight-point MHV amplitude up to three loops~\cite{Caron-Huot:2011zgw,Bargheer:2015djt,DelDuca:2018raq,He:2019jee,Li:2021bwg}. We perform the analytic continuation of the known two- and three-loop expressions for the octagon to the zigzag region. The analytic continuation can be carried out using the coproduct~\cite{Goncharov:2001iea,Goncharov:2005sla,brownmixedZ,Duhr:2012fh} of $R_{8}^{(L)}$ following the procedure outlined in ref.~\cite{DelDuca:2018raq}. After analytic continuation we take the multi-Regge limit. Further details of this procedure for the octagon $R_{8}^{(L)}$ are provided in appendix~\ref{app:A}.  We now discuss the results at two and three loops in detail.

\paragraph*{Two loops.} First of all, we find from the perturbative data in ref.~\cite{Caron-Huot:2011zgw} (or more directly from ref.~\cite{DelDuca:2018raq}) that
\begin{align}
    \mathcal{R}_{\{4,7\},8}\me^{\pi\mi\tilde{a}\delta_{8}^{\{4,7\}}} \Big\vert_{\text{2-loop}}
    &=2\pi\mi X_{\{4,7\},8}^{(2),1} - \frac{(2\pi\mi)^2}{2}\log\frac{|\mathbf{x}_{23}\mathbf{x}_{56}|^2}{|\mathbf{x}_{25}\mathbf{x}_{36}|^2}\log\frac{|\mathbf{x}_{26}\mathbf{x}_{35}|^2}{|\mathbf{x}_{25}\mathbf{x}_{36}|^2}\nonumber \\
    &\quad+\frac{(2\pi\mi)^2}{8}\ab(\delta_{8}^{\{4,7\}})^2\,.
\end{align}
Here, $X_{\{4,7\},8}^{(2),1}$ can be obtained by combining eq.~\eqref{eq:Xrel1} with the known result of
ref.~\cite{Bargheer:2015djt},
\begin{equation*}
         X_{[j,k],N}^{(2),1}=\sum_{i=j}^{k-1}\ab[2\log(\tau_{i})f(\rho_{i})+\tilde{f}(\rho_{i})]
         +\sum_{i=j}^{k-2}g(\rho_{i},\rho_{i+1}) \:,
\end{equation*}
where $\tilde{f}(\rho_i)$ and $g(\rho_i,\rho_{i+1})$ are weight-three polylogarithms whose explicit expressions are not relevant for our purposes, while $f(\rho)$ is given by~\cite{Lipatov:2010ad,Bartels:2011ge,Prygarin:2011gd}
\begin{equation}
    \label{eq: two_loop_ffunction}
    f(\rho)
    =
    \frac{1}{2}\log|1-\rho|^2
    \log\ab|1-\rho^{-1}|^2 \,.
\end{equation}
On the other hand, eq.~\eqref{eq:two_reggeon_sub}, together with the fact that $\mathcal{R}_{[j,k],N}^{(2)}=(2\pi\mi)X_{[j,k],N}^{(2),1}$ (see ref.~\cite{DelDuca:2018raq}), gives
\begin{align}
    &\mathcal{W}_{\{4,7\}}^{(2)}=  X_{\{4,7\},8}^{(2),1} - 2\pi\mi
    \ab[f\ab(\frac{\mathbf{x}_{32}\mathbf{x}_{16}}{\mathbf{x}_{12}\mathbf{x}_{36}})+f\ab(\frac{\mathbf{x}_{52}\mathbf{x}_{16}}{\mathbf{x}_{12}\mathbf{x}_{56}})-f\ab(\frac{\mathbf{x}_{32}\mathbf{x}_{15}}{\mathbf{x}_{12}\mathbf{x}_{35}})-f\ab(\frac{\mathbf{x}_{53}\mathbf{x}_{16}}{\mathbf{x}_{13}\mathbf{x}_{56}})] \\
    &\quad +\frac{2\pi\mi}{8}\ab[\ab(\delta_{8}^{[4,7]})^2-\ab(\delta_{8}^{[4,6]}-\log|z_{3}|^2)^2-\ab(\delta_{8}^{[5,7]}+\log|z_{1}|^2)^2+\ab(\delta_{8}^{[5,6]}+\log\ab|\frac{z_{1}}{z_{3}}|^2)^2] \:,  \nonumber
\end{align}
where the second term on the right-hand side arises from the replacement $\log\tau_{1,3}\to\log\tau_{1,3}-\pi\mi$. We then obtain from eq.~\eqref{eq:dec_oct} that
\begin{align} \label{eq: three_reg_perturbative_2loop}
    \mathcal{T}_{\{4,7\}}^{(2)}=\frac{1}{2} \log\ab|\frac{\mathbf{x}_{25}\mathbf{x}_{36}}{\mathbf{x}_{26}\mathbf{x}_{35}}|^2 \log \ab| \frac{\mathbf{x}_{12}\mathbf{x}_{16}\mathbf{x}_{23}\mathbf{x}_{56}}{\mathbf{x}_{13}\mathbf{x}_{15}\mathbf{x}_{26}^2}|^2 \:.
\end{align}
Note that $\mathcal{T}_{\{4,7\}}^{(2)}=\delta_{8}^{\{4,7\}}\delta_{8}^{[4,7]}/4$. In the next subsection we will present the computation of $\mathcal{T}_{\{4,7\}}^{(2)}$ directly from the BFKL-EFT, and we see that it agrees with eq.~\eqref{eq: three_reg_perturbative_2loop}. This shows that our proposal in eq.~\eqref{eq:two_reggeon_sub} is consistent with the available perturbative data at two loops.

\paragraph*{Three loops.} Similarly, all the quantities required at three loops, including $X_{\{4,7\},8}^{(3),1}$ and $X_{\{4,7\},8}^{(3),2}$, can be extracted from the known three-loop octagon~\cite{Li:2021bwg} as shown in appendix~\ref{app:A}. We perform the analytic continuation, take the multi-Regge limit, and subtract the contribution from two-Reggeon exchange through NLLA obtained from eq.~\eqref{eq:2Reggeon_47} (see ref.~\cite{Marzucca:2018ydt}), and we obtain the following expression for $\mathcal{T}_{\{4,7\}}^{(3)}$:
\begin{equation} \label{eq: three_reg_perturbative_3loop}
  \mathcal{T}_{\{4,7\}}^{(3)} = 
   -\sum_{i=1}^{3} \log(\tau_{i})  \tilde{h}_{i}(\rho_{1},\rho_{2},\rho_{3})
   +\kappa^{(4)}(\rho_{1},\rho_{2},\rho_{3})\,,
\end{equation}
where $\kappa^{(4)}$ is a weight-four single-valued polylogarithm whose explicit form is not needed here, but it is provided in the ancillary file \texttt{MHV3Reggeon.m}. The coefficients of $\log\tau_{i}$, which we will reproduce in the next subsection, can be written compactly as
\begin{align}
 \tilde{h}_{2}(\rho_{1},\rho_{2},\rho_{3})&=-\calG_{1,0,\rho_{3}}(\rho_{1})
- \calG_{1,\rho_{3},1}(\rho_{1})
- \calG_{\rho_{3},0,1}(\rho_{1})  \nonumber \\
&\quad
+ \ab[\frac{1}{2}\log\ab|1-\rho_{1}|^{2}
\log\ab|\frac{\rho_{1}}{\rho_{3}}|^{2}+\calG_{\rho_{3},1}(\rho_{1})]
\log\ab|\frac{\rho_{3}}{1-\rho_{3}}|^{2} \nonumber \\
&\quad
- \frac{1}{2}\log\ab|1-\frac{\rho_{1}}{\rho_{3}}|^{2}
\Bigl[
 \log\ab|\rho_{3}|^{2}\log\ab|(1-\rho_{3})\rho_{1}(1-\rho_{1})|^{2}
 -\log^{2}\ab|\rho_{1}|^{2}
- 2\calG_{0,1}(\rho_{3})
\Bigr] \nonumber \\
&\quad
- \frac{1}{2}\log\ab|\frac{(1-\rho_{2})^{2}\rho_{1}}{\rho_{2}}|^{2}
\log\ab|1-\frac{\rho_{1}}{\rho_{3}}|^{2}
\log\ab|\frac{\rho_{1}}{(1-\rho_{1})(1-\rho_{3})}|^{2}  \nonumber \\
&\quad
+ \frac{1}{2}\log\ab|\frac{\rho_{1}-\rho_{3}}{\rho_{2}-\rho_{3}}|^{2}
\log\ab|\frac{\rho_{1}}{(1-\rho_{1})(1-\rho_{3})}|^{2}
\log\ab|\frac{(\rho_{1}-\rho_{2})\rho_{3}}{\rho_{2}(\rho_{1}-\rho_{3})}|^{2} \:, \label{eq: htilde2}
\end{align}
where $\calG_{\vec{a}}(z):=\calG(\vec{a};z)$ denotes a single-valued polylogarithm (for a detailed discussion of this class of functions, see ref.~\cite{DelDuca:2016lad} and the references therein).
The remaining two coefficients $\tilde{h}_{1,3}$ can be obtained from $\tilde{h}_{2}$ by taking $\rho_{2}\to \rho_{1,3}$. Furthermore, the above expression for $\tilde{h}_{2}$ makes manifest the fact that all $\tilde{h}_{i}$ vanish as $\rho_{1}\to 0$ or $\rho_{3}\to \infty$, which agrees with the corresponding soft behaviors of $\mathcal{R}_{\{4,7\},8}$. In the next subsection we will perform the explicit computation of $  \mathcal{T}_{\{4,7\}}^{(3)}$ directly from the BFKL-EFT. We find perfect agreement with eqs.~\eqref{eq: three_reg_perturbative_3loop} and~\eqref{eq: htilde2}, confirming that our proposal in eq.~\eqref{eq:two_reggeon_sub} is consistent with the available perturbative data also at three loops.

\subsection{Three-Reggeon exchange from an EFT computation}

In this subsection, we describe how to compute the three-Reggeon exchange contribution to the zigzag octagon. This serves two purposes: first, it provides a check of the proposal in eq.~\eqref{eq:two_reggeon_sub} given the decomposition in eq.~\eqref{eq:dec_oct}; second, it predicts the three-Reggeon exchange contributions to
as-yet-unknown amplitudes, such as higher-point amplitudes at three loops and the octagon at four loops.

At two loops, only one Feynman diagram contributes to $\mathcal{T}_{\{4,7\}}^{(2)}$:
\begin{align} \label{eq: MHV_Three_Reggeon}
  \begin{tikzpicture}[baseline={([yshift=-0.6ex]current bounding box.center)}]
        \draw[line width=0.3mm] (1,1.0)--(-1,1.0) (1,-1.0)--(-1,-1.0);
        \draw[line width=0.3mm] (0.5,1.0)node[circle,draw=black, fill=black, inner sep=0.2ex]{} -- (0.5,-1.0)node[circle,draw=black, fill=black, inner sep=0.2ex]{} 
        (-0.5,1.0)node[circle,draw=black, fill=black, inner sep=0.2ex]{} -- (-0.5,-1.0)node[circle,draw=black, fill=black, inner sep=0.2ex]{};
        \draw[line width=0.3mm] (0.5,0.5)node[circle,draw=black, fill=black, inner sep=0.2ex]{} -- (1.5,0.5)
        (0.5,-0.5)node[circle,draw=black, fill=black, inner sep=0.2ex]{} -- (1.5,-0.5);
         \draw[line width=0.3mm] (1.2,0.5)node[circle,draw=black, fill=black, inner sep=0.2ex]{} -- (1.2,-0.5)node[circle,draw=black, fill=black, inner sep=0.2ex]{} ;
         \draw[line width=0.3mm] (1.2,0.2)node[circle,draw=black, fill=black, inner sep=0.2ex]{} -- (1.5,0.2)
        (1.2,-0.2)node[circle,draw=black, fill=black, inner sep=0.2ex]{} -- (1.5,-0.2);
        \node at (0,0) {$\mathbf{x}_{a}$};
        \node at (0.9,0) {$\mathbf{x}_{b}$};
    \end{tikzpicture}&=\int\frac{\mathbf{x}_{12}\mathbf{x}^{\ast}_{16}\:\dbar^2\mathbf{x}_{a}}{|\mathbf{x}_{1a}|^2 \mathbf{x}_{a2}\mathbf{x}_{a6}^{\ast}} \frac{\mathbf{x}_{23}\mathbf{x}^{\ast}_{56}\dbar^2\mathbf{x}_{b}}{\mathbf{x}_{2b}\mathbf{x}_{b3}\mathbf{x}^{\ast}_{5b}\mathbf{x}^{\ast}_{b6}} = \log\frac{|\mathbf{x}_{12}|^{2}|\mathbf{x}_{16}|^{2}}{|\mathbf{x}_{26}|^2\mu^2} \log\ab|\frac{\mathbf{x}_{25}\mathbf{x}_{36}}{\mathbf{x}_{26}\mathbf{x}_{35}}|^2  \:.
\end{align}
This two-loop integral factorizes, but it is unfortunately divergent due to the integration over $\mathbf{x}_a$. However, this divergence is of the same kind as that encountered in the one-loop integral in eq.~\eqref{eq:six_oneloop} for the two-Reggeon exchange. We therefore follow the same prescription and remove the divergence by subtracting suitable triangle integrals, which amounts to making the following replacement for the octagon case:
\begin{equation} \label{eq: subtraction_uni}
    \mu^2 \to \frac{|\mathbf{x}_{12}\mathbf{x}_{13}\mathbf{x}_{15}\mathbf{x}_{16}|}{|\mathbf{x}_{23}\mathbf{x}_{56}|} \:.
\end{equation}
With this replacement, the integrated result in eq.~\eqref{eq: MHV_Three_Reggeon} is then in perfect agreement with the perturbative data in eq.~\eqref{eq: three_reg_perturbative_2loop}.

For the $(L+2)$-loop contribution from the three-Reggeon exchange at LLA, the loop insertions are organized in the same way as in the two-Reggeon case; explicitly, this is
\begin{align}
     \begin{tikzpicture}[baseline={([yshift=-0.6ex]current bounding box.center)}]
           \draw[line width=0.3mm]  (0.45,-0.7)--(0.45,0.7) (0,0.7)--(0,-0.7) (-0.45,-0.7)--(-0.45,0.7);
          \node[rectangle, draw = black, fill = gray!40,font=\footnotesize,inner sep=1.5pt,minimum width=1.3cm,minimum height=0.7cm] at (0,0) {$\widehat{G}_{3}^{(L)}$};
          \node[right,font=\footnotesize] at (0.6,-0.05) {$\mathbf{x}_{j}$};
          \node[left,font=\footnotesize] at (-0.6,-0.05) {$\mathbf{x}_{i}$};
    \end{tikzpicture} &=  \hspace{0.5em}
    \begin{tikzpicture}[baseline={([yshift=-0.6ex]current bounding box.center)}]
           \fill[black] (0,0.55) circle (0.3ex);
           \fill[black] (-0.45,0.55) circle (0.3ex);
           \draw[line width=0.3mm]  (0.45,-0.5)--(0.45,0.9) (0,-0.5)--(0,0.9) (-0.45,-0.5)--(-0.45,0.9) (0,0.55)--(-0.45,0.55);
          \node[rectangle, draw = black, fill = gray!40,font=\footnotesize,inner sep=1.5pt,minimum width=1.3cm] at (0,0) {$\widehat{G}_{3}^{(L-1)}$};
    \end{tikzpicture}   \hspace{0.5em}  + \hspace{0.5em}
    \begin{tikzpicture}[baseline={([yshift=-0.6ex]current bounding box.center)}]
           \fill[black] (0,0.55) circle (0.3ex);
           \fill[black] (0.45,0.55) circle (0.3ex);
           \draw[line width=0.3mm]  (0.45,-0.5)--(0.45,0.9) (0,-0.5)--(0,0.9) (-0.45,-0.5)--(-0.45,0.9) (0,0.55)--(0.45,0.55);
          \node[rectangle, draw = black, fill = gray!40,font=\footnotesize,inner sep=1.5pt,minimum width=1.3cm] at (0,0) {$\widehat{G}_{3}^{(L-1)}$};
    \end{tikzpicture}   \hspace{0.5em}
    -   \hspace{0.5em} 
    \begin{tikzpicture}[baseline={([yshift=-0.6ex]current bounding box.center)}]
           \draw[line width=0.3mm]  (0.45,-0.5)--(0.45,0.9) (0,-0.5)--(0,0.9) (-0.45,-0.5)--(-0.45,0.9);
            \draw [black,fill=white,line width=0.2mm] (-0.45,0.55) ellipse (0.3em and 0.4em);
          \node[rectangle, draw = black, fill = gray!40,font=\footnotesize,inner sep=1.5pt,minimum width=1.3cm] at (0,0) {$\widehat{G}_{3}^{(L-1)}$};
    \end{tikzpicture}  \hspace{0.5em}
    -   \hspace{0.5em}
       \begin{tikzpicture}[baseline={([yshift=-0.6ex]current bounding box.center)}]
           \draw[line width=0.3mm]  (0.45,-0.5)--(0.45,0.9) (0,-0.5)--(0,0.9) (-0.45,-0.5)--(-0.45,0.9);
            \draw [black,fill=white,line width=0.2mm] (0,0.55) ellipse (0.3em and 0.4em);
          \node[rectangle, draw = black, fill = gray!40,font=\footnotesize,inner sep=1.5pt,minimum width=1.3cm] at (0,0) {$\widehat{G}_{3}^{(L-1)}$};
    \end{tikzpicture}   \hspace{0.5em} 
    -  \hspace{0.5em} 
        \begin{tikzpicture}[baseline={([yshift=-0.6ex]current bounding box.center)}]
           \draw[line width=0.3mm]  (0.45,-0.5)--(0.45,0.9) (0,-0.5)--(0,0.9) (-0.45,-0.5)--(-0.45,0.9);
            \draw [black,fill=white,line width=0.2mm] (0.45,0.55) ellipse (0.3em and 0.4em);
          \node[rectangle, draw = black, fill = gray!40,font=\footnotesize,inner sep=1.5pt,minimum width=1.3cm] at (0,0) {$\widehat{G}_{3}^{(L-1)}$};
    \end{tikzpicture}  
    \nonumber \\
     &\quad  + 2\log\frac{|\mathbf{x}_{ij}|^2}{\mu^2}
    \ab(\begin{tikzpicture}[baseline={([yshift=-0.6ex]current bounding box.center)}]
       \draw[line width=0.3mm]  (0.45,-0.5)--(0.45,0.9) (0,-0.5)--(0,0.9) (-0.45,-0.5)--(-0.45,0.9);
          \node[rectangle, draw = black, fill = gray!40,font=\footnotesize,inner sep=1.5pt,minimum width=1.3cm] at (0,0) {$\widehat{G}_{3}^{(L-1)}$};
    \end{tikzpicture})
     \:.
        \label{eq: three_BFKL_bridge}
\end{align}
The construction for the general multi-Reggeon exchange follows through a straightforward generalization.

To obtain the coefficient of $\log\tau_{i}$ in $\mathcal{T}_{\{4,7\}}^{(3)}$, one can simply insert $\widehat{G}_{3}^{(1)}$ into the rapidity region $[\eta_{i+3},\eta_{i+4}]$. Unlike the case of the two-Reggeon exchange, where the multi-loop contributions are free of IR divergences due to the insertion of $\widehat{G}_{A}^{(L)}$, the multi-loop contributions from the three-Reggeon exchange and beyond are still divergent, but they take a rather simple form and can be removed through the same prescription as above. For example, the divergent piece appearing in the Feynman diagram computation for $\tilde{h}_{2}$ is of the form
\begin{equation}
     2\log(\mu^2)\ab[f\ab(\frac{\mathbf{x}_{42}\mathbf{x}_{16}}{\mathbf{x}_{12}\mathbf{x}_{46}})+f\ab(\frac{\mathbf{x}_{43}\mathbf{x}_{15}}{\mathbf{x}_{13}\mathbf{x}_{45}})-f\ab(\frac{\mathbf{x}_{42}\mathbf{x}_{15}}{\mathbf{x}_{12}\mathbf{x}_{45}})-f\ab(\frac{\mathbf{x}_{43}\mathbf{x}_{16}}{\mathbf{x}_{13}\mathbf{x}_{46}})] \:,
\end{equation}
which is nothing but the coefficient of $\log \tau_{2}$ in $X_{\{4,7\},8}^{(2),1}$ up to an overall numerical constant. Removing this divergence using the replacement in eq.~\eqref{eq: subtraction_uni} then reproduces the result for $\tilde{h}_{2}$ in eq.~\eqref{eq: htilde2}. The results for $\tilde{h}_{1}$ and $\tilde{h}_{3}$ are also reproduced simultaneously, since their corresponding Feynman integrands coincide with the $\mathbf{x}_4\to\mathbf{x}_3$ and $\mathbf{x}_4\to\mathbf{x}_5$ limits of the integrands for $\tilde{h}_2$, respectively.

Higher-loop results can be obtained in the same way, except that now there are multiple ways of inserting $\widehat{G}_{3}^{(L)}$. For example, at four loops there are six distinct insertions, each yielding the coefficient of $\log\tau_{i}\log\tau_{j}$ with $i\leq j$ in $\mathcal{T}_{\{4,7\}}^{(4)}$. The IR divergence is of the same type as that encountered at two and three loops. This is because it arises only from diagrams containing the right reduced BFKL ladders, in which the divergence associated with the leftmost propagator $1/|\mathbf{x}_{1a}|^2$ is not suppressed. Therefore, the IR divergence in $\mathcal{T}_{\{4,7\}}$ is well controlled and given by
\begin{equation}
 \begin{tikzpicture}[baseline={([yshift=-0.6ex]current bounding box.center)}]
        \draw[line width=0.3mm] (1,1.5)--(-1,1.5) (1,-1.5)--(-1,-1.5);
        \draw[line width=0.3mm] (0.5,1.5)node[circle,draw=black, fill=black, inner sep=0.2ex]{} -- (0.5,-1.5)node[circle,draw=black, fill=black, inner sep=0.2ex]{} 
        (-0.5,1.5)node[circle,draw=black, fill=black, inner sep=0.2ex]{} -- (-0.5,-1.5)node[circle,draw=black, fill=black, inner sep=0.2ex]{};
        \draw[line width=0.3mm] 
        (0.5,1.2)node[circle,draw=black, fill=black, inner sep=0.2ex]{} -- (1.5,1.2)
        (0.5,-1.2)node[circle,draw=black, fill=black, inner sep=0.2ex]{} -- (1.5,-1.2);
         \draw[line width=0.3mm] 
         (1.2,1.2)node[circle,draw=black, fill=black, inner sep=0.2ex]{} -- (1.2,-1.2)node[circle,draw=black, fill=black, inner sep=0.2ex]{} ;
         \draw[line width=0.3mm] (1.2,0.45)node[circle,draw=black, fill=black, inner sep=0.2ex]{} -- (1.5,0.45)
        (1.2,-0.45)node[circle,draw=black, fill=black, inner sep=0.2ex]{} -- (1.5,-0.45);
        \node[rectangle, draw = black, fill = gray!40,font=\footnotesize,inner sep=1.5pt,minimum width=1.0cm,minimum height=0.4cm] at (0.85,0) {$\widehat{G}_{A}$};
         \node[rectangle, draw = black, fill = gray!40,font=\footnotesize,inner sep=1.5pt,minimum width=1.0cm,minimum height=0.4cm] at (0.85,-0.82) {$\widehat{G}_{A}$};
          \node[rectangle, draw = black, fill = gray!40,font=\footnotesize,inner sep=1.5pt,minimum width=1.0cm,minimum height=0.4cm] at (0.85,0.82) {$\widehat{G}_{A}$};
    \end{tikzpicture} \quad \propto \quad \log\mu^2 \times \mathcal{W}_{\{4,7\}}\big|_{\text{LL}} \:.
\end{equation}
This justifies the IR subtraction prescription \eqref{eq: subtraction_uni} to all orders. We have computed these coefficients, and the results are collected in the file \texttt{MHV3Reggeon.m}. We stress that these are completely new results for which there are currently no existing perturbative data to test against. Note that we could also evaluate the contributions from two-Reggeon exchange at four loops, $\mathcal{W}_{\{4,7\}}^{(4)}$, from eq.~\eqref{eq:two_reggeon_sub}. By combining them as in eq.~\eqref{eq:dec_oct} we obtain for the first time the complete result for the BDS-normalized eight-point amplitude in MRK in the Mandelstam region $\{4,7\}$.

The generalization to higher points is likewise straightforward, in analogy with the two-Reggeon exchange case: the $N$-point amplitude contains a three-Reggeon exchange in the Mandelstam region $\{4,N-1\}$. First, the same reasoning shows that the two-Reggeon contribution $\mathcal{W}_{\{4,7\}}$ generalizes directly to
\begin{equation}\label{eq:W_{4,N-1}}
    \mathcal{W}_{\{4,N-1\}}= g^2 \prod_{r=1}^{N-5}\ab[\sum_{n_r}\ab(\frac{z_{r}}{\bar{z}_{r}})^{n_{r}/2}\int_{\mathcal{C}'}\frac{\dif \nu_{r}}{2\pi}]  \frac{  \chi^{\oplus}_{1}\,C^{\oplus}_{12} \cdots C^{\oplus}_{(N-6)(N-5)}\,\chi^{\ominus}_{N-5}}{\tau_{1}^{\omega_1}(-\tau_{2}+\mi 0)^{\omega_2}\cdots (-\tau_{N-6}+\mi 0)^{\omega_{N-6}}\tau_{N-5}^{\omega_{N-5}}} \:,
\end{equation}
where again the integration contour $\mathcal{C}'$ for $\nu_{1}$ and $\nu_{N-5}$ is deformed as in figure~\ref{fig:deformed_contour}.
Second, for the three-Reggeon exchange at three loops, by examining the integrand of the corresponding Feynman diagram, one finds that the Feynman integrals are of the same type as those in the octagon case, but now the IR divergence should be removed through the replacement
\begin{equation}
  \mu^{2}\to \frac{|\mathbf{x}_{1,2}\mathbf{x}_{1,3}\mathbf{x}_{1,N-3}\mathbf{x}_{1,N-2}|}{|\mathbf{x}_{2,3}\mathbf{x}_{N-3,N-2}|}  \:.
\end{equation}
The result can therefore be expressed as
\begin{equation}
  \mathcal{T}_{\{4,N-1\}}^{(3)} = -\sum_{i=1}^{N-5} \log \tau_{i}\, \tilde{h}_{i}(\rho_{1},\rho_{i},\rho_{N-5}) +\cdots\:,
\end{equation}
where the dots indicate terms proportional to $2\pi\mi$ and $\tilde{h}_{i}$ is the same function as $\tilde{h}_{2}$ introduced in eq.~\eqref{eq: htilde2}. We see that, for any number of points, the contribution from three-Reggeon exchange at three loops at LLA is entirely determined by the eight-point result. This is very similar to the factorization observed for two-Reggeon exchange in refs.~\cite{Prygarin:2011gd,Bartels:2011ge}, where the LLA result is entirely determined by the behavior of the six-point amplitude. Combining this with the result for two-Reggeon exchange in eq.~\eqref{eq:W_{4,N-1}}, we obtain predictions for all MHV amplitudes at NLLA for arbitrary $N$ in the regions $\{4,N-1\}$.

\subsection{Results for non-MHV amplitudes}

In this subsection, we describe the computation of the multi-Regge limit of non-MHV octagons in the Mandelstam region $\{4,7\}$. We denote them by $\mathcal{R}_{\varrho}^{h_4h_5h_6h_7} := \mathcal{R}_{\varrho,8}^{h_4h_5h_6h_7}$, where $h_i$ denotes the helicity of produced particle $i$. In this notation, the MHV case simply corresponds to setting $h_i=\oplus$ for all produced particles.

As in the MHV case, the contributions to $\mathcal{R}_{\{4,7\}}^{h_{4}h_{5}h_{6}h_{7}} \me^{\pi\mi\tilde{a}\delta_{8}^{\{4,7\}}}$ arise from two-Reggeon exchange and three-Reggeon exchange, and hence can be decomposed as (cf.~eq.~\eqref{eq:dec_oct})
\begin{equation} \label{eq: nonMHVocta_MRK}
    \mathcal{R}_{\{4,7\}}^{h_{4}h_{5}h_{6}h_{7}}\me^{\pi\mi\tilde{a}\delta_{8}^{\{4,7\}}} 
    =1+ (2\pi\mi) \mathcal{W}_{\{4,7\}}^{h_{4}h_{5}h_{6}h_{7}} 
    + (2\pi\mi)^2 \mathcal{T}_{\{4,7\}}^{h_{4}h_{5}h_{6}h_{7}} \:,
\end{equation}
where, in analogy with eq.\ \eqref{eq:two_reggeon_sub}, the two-Reggeon exchange for the non-MHV case can be written as
\begin{align} \label{eq: nonMHV_two_reggeon_sub}
     \mathcal{W}_{\{4,7\}}^{h_{4}h_{5}h_{6}h_{7}} &=  \Bigl[
     \mathcal{W}^{h_{4}h_{5}h_{6}h_{7}}_{[4,7]}
     -\mathcal{W}^{h_{4}h_{5}h_{6}}_{[4,6]}|z_{3}|^{-2\pi\mi\tilde{a}}
     -\mathcal{W}_{[5,7]}^{h_{5}h_{6}h_{7}}|z_{1}|^{2\pi\mi\tilde{a}}  
     \nonumber \\ 
     &\quad +\mathcal{W}_{[5,6]}^{h_{5}h_{6}} |z_{1}/z_{3}|^{2\pi\mi\tilde{a}}\Bigr]
     \Big\vert_{\log\tau_{1,3}\to\log\tau_{1,3}-\mi\pi}\:.
\end{align}
Here, all BDS-subtracted amplitudes are understood to be normalized by the tree-level amplitudes with the same helicity configuration.
A few remarks are in order:
\begin{itemize}
    \item In contrast to the MHV case, the non-MHV BDS-subtracted amplitudes receive one-loop corrections, which are entirely determined by the two-Reggeon exchange.
    \item The result for $\mathcal{W}^{h_{4}h_{5}h_{6}h_{7}}_{[4,7]}$ can be found in ref.~\cite{Marzucca:2018ydt} up to three-loop order; the other terms on the right-hand side of eq.\ \eqref{eq: nonMHV_two_reggeon_sub} can be obtained through various soft limits.
    \item The three-Reggeon exchange starts contributing at two loops, as in the MHV case.
    \item We only need to consider five different helicity configurations: ${\ominus}{\oplus}{\oplus}{\oplus}$, ${\oplus}{\ominus}{\oplus}{\oplus}$, ${\ominus}{\oplus}{\ominus}{\oplus}$, ${\oplus}{\ominus}{\ominus}{\oplus}$, and ${\ominus}{\ominus}{\oplus}{\oplus}$. The remaining helicity configurations can be obtained by using projectile-target symmetry, which exchanges $h_{i}$ and $h_{N+3-i}$, and/or taking the complex conjugate, which replaces $h_{i}$ with $-h_{i}$.\footnote{Due to the different conventions for the polarization vectors, the results we obtained for the non-MHV cases are the complex conjugates of those in refs.~\cite{DelDuca:2016lad,DelDuca:2018hrv,Marzucca:2018ydt,DelDuca:2019tur,Dixon:2021nzr}. Therefore, we express these results in terms of the complex-conjugate variables, such as $\bar{\rho}_{i}$ and $\mathbf{x}_{i}^{\ast}$, for comparison. The MHV cases are not affected, since the result is real.}
\end{itemize}

\paragraph{One loop.} To obtain $\mathcal{R}_{\{4,7\}}^{h_{4}h_{5}h_{6}h_{7}}$ at one loop, we only need the result for $\mathcal{R}_{[4,7]}^{(1),h_{4}h_{5}h_{6}h_{7}}$, since eqs.~\eqref{eq: nonMHVocta_MRK} and \eqref{eq: nonMHV_two_reggeon_sub} at one loop reduce to
\begin{equation}
    \mathcal{R}_{\{4,7\}}^{(1),h_{4}h_{5}h_{6}h_{7}}=  \mathcal{R}_{[4,7]}^{(1),h_{4}h_{5}h_{6}h_{7}}
    - \mathcal{R}_{[5,7]}^{(1),h_{5}h_{6}h_{7}}- \mathcal{R}_{[4,6]}^{(1),h_{4}h_{5}h_{6}}
    + \mathcal{R}_{[5,6]}^{(1),h_{5}h_{6}} \:,
\end{equation}
and the last three terms on the right-hand side are simply various soft limits of $\mathcal{R}_{[4,7]}^{(1),h_{4}h_{5}h_{6}h_{7}}$.
The one-loop result can most easily be obtained through the EFT. For example, the one-loop BDS-subtracted amplitude $\mathcal{R}^{(1),\ominus\oplus\oplus\oplus}_{[4,7]}$ is given by
\begin{align}
    \frac{\mathcal{R}^{(1),\ominus\oplus\oplus\oplus}_{[4,7]}}{2\pi\mi} &= \begin{tikzpicture}[baseline={([yshift=-0.6ex]current bounding box.center)}]
        \draw[line width=0.3mm] (1,1.5)--(-1,1.5) (1,-1.5)--(-1,-1.5);
        \draw[line width=0.3mm] (0.5,1.5)node[circle,draw=black, fill=black, inner sep=0.2ex]{} -- (0.5,-1.5)node[circle,draw=black, fill=black, inner sep=0.2ex]{} 
        (-0.5,1.5)node[circle,draw=black, fill=black, inner sep=0.2ex]{} -- (-0.5,-1.5)node[circle,draw=black, fill=black, inner sep=0.2ex]{};
        \draw[line width=0.3mm] (0.5,1.0)node[circle,draw=black, fill=black, inner sep=0.2ex]{} -- (1.0,1.0)node[right]{$\ominus$}
        (0.5,-1.0)node[circle,draw=black, fill=black, inner sep=0.2ex]{} -- (1.0,-1.0)node[right]{$\oplus$};
        \draw[line width=0.3mm] (0.5,0.33)node[circle,draw=black, fill=black, inner sep=0.2ex]{} -- (1.0,0.33)node[right]{$\oplus$}
        (0.5,-0.33)node[circle,draw=black, fill=black, inner sep=0.2ex]{} -- (1.0,-0.33)node[right]{$\oplus$};
    \end{tikzpicture} \quad - \quad 
    \begin{tikzpicture}[baseline={([yshift=-0.6ex]current bounding box.center)}]
        \draw[line width=0.3mm] (1,1.5)--(-1,1.5) (1,-1.5)--(-1,-1.5);
        \draw[line width=0.3mm] (0.5,1.5)node[circle,draw=black, fill=black, inner sep=0.2ex]{} -- (0.5,-1.5)node[circle,draw=black, fill=black, inner sep=0.2ex]{} 
        (-0.5,1.5)node[circle,draw=black, fill=black, inner sep=0.2ex]{} -- (-0.5,-1.5)node[circle,draw=black, fill=black, inner sep=0.2ex]{};
        \draw[line width=0.3mm] (0.5,1.0)node[circle,draw=black, fill=black, inner sep=0.2ex]{} -- (1.0,1.0)node[right]{$\oplus$}
        (0.5,-1.0)node[circle,draw=black, fill=black, inner sep=0.2ex]{} -- (1.0,-1.0)node[right]{$\oplus$};
        \draw[line width=0.3mm] (0.5,0.33)node[circle,draw=black, fill=black, inner sep=0.2ex]{} -- (1.0,0.33)node[right]{$\oplus$}
        (0.5,-0.33)node[circle,draw=black, fill=black, inner sep=0.2ex]{} -- (1.0,-0.33)node[right]{$\oplus$};
    \end{tikzpicture} \nonumber \\[1em]
    &=\int \ab(\frac{\mathbf{x}_{13}\mathbf{x}^{\ast}_{12}\mathbf{x}_{16}^{\ast}\mathbf{x}_{a3}^{\ast}}{\mathbf{x}_{a1}\mathbf{x}_{a3}\mathbf{x}^{\ast}_{13}\mathbf{x}^{\ast}_{a1}\mathbf{x}^{\ast}_{a2}\mathbf{x}_{a6}} - \frac{\mathbf{x}_{12}\mathbf{x}^{\ast}_{16}}{|\mathbf{x}_{a1}|^2\mathbf{x}_{a2}\mathbf{x}^{\ast}_{a6}}) \dbar^2 \mathbf{x}_{a} \nonumber \\
    &=\bar{R}_{236} \log \ab|\frac{\mathbf{x}_{12}\mathbf{x}_{36}}{\mathbf{x}_{23}\mathbf{x}_{16}}|^2 - \log \ab|\frac{\mathbf{x}_{12}\mathbf{x}_{36}}{\mathbf{x}_{13}\mathbf{x}_{26}}|^2 \:,
\end{align}
where we have introduced, as in ref.~\cite{DelDuca:2016lad},
\begin{equation}
    R_{bac}=\frac{\mathbf{x}_{ba}\mathbf{x}_{c1}}{\mathbf{x}_{bc}\mathbf{x}_{a1}}\,,
\end{equation}
as well as its complex conjugate $\bar{R}_{bac}$. Note that the IR divergences of individual diagrams cancel out in the difference, and $\mathcal{R}^{(1),\ominus\oplus\oplus\oplus}_{[4,7]}$ vanishes in the soft limit $\mathbf{x}_{3}\to \mathbf{x}_{2}$, as expected for $\mathcal{R}^{(1),\oplus\oplus\oplus}_{[5,7]}$.
Similarly, we obtain
\begingroup
\allowdisplaybreaks
\begin{align}
    \frac{\mathcal{R}^{(1),\oplus\ominus\oplus\oplus}_{[4,7]}}{2\pi\mi} &=  (R_{234}-1) \log \ab|\frac{\mathbf{x}_{12}\mathbf{x}_{46}}{\mathbf{x}_{14}\mathbf{x}_{26}}|^2 + \bar{R}_{346}\log\ab|\frac{\mathbf{x}_{13}\mathbf{x}_{46}}{\mathbf{x}_{34}\mathbf{x}_{16}}|^2  + R_{234}\bar{R}_{346}\log\ab|\frac{\mathbf{x}_{34}\mathbf{x}_{26}}{\mathbf{x}_{23}\mathbf{x}_{46}}|^2 \:, \\ 
    \frac{\mathcal{R}^{(1),\ominus\ominus\oplus\oplus}_{[4,7]}}{2\pi\mi} &= \log\ab|\frac{\mathbf{x}_{14}\mathbf{x}_{26}}{\mathbf{x}_{12}\mathbf{x}_{46}}|^2 + \bar{R}_{246} \log\ab|\frac{\mathbf{x}_{12}\mathbf{x}_{46}}{\mathbf{x}_{24}\mathbf{x}_{16}}|^2  \:, \\
    \frac{\mathcal{R}^{(1),\oplus\ominus\ominus\oplus}_{[4,7]}}{2\pi\mi}&= (R_{235}-1)\log\ab|\frac{\mathbf{x}_{12}\mathbf{x}_{56}}{\mathbf{x}_{15}\mathbf{x}_{26}}|^2+\bar{R}_{356}\log\ab|\frac{\mathbf{x}_{13}\mathbf{x}_{56}}{\mathbf{x}_{35}\mathbf{x}_{16}}|^2 +R_{235}\bar{R}_{356}\log\ab|\frac{\mathbf{x}_{35}\mathbf{x}_{26}}{\mathbf{x}_{23}\mathbf{x}_{56}}|^2  \:,\\ 
    \frac{\mathcal{R}^{(1),\ominus\oplus\ominus\oplus}_{[4,7]}}{2\pi\mi}&= \log\ab|\frac{\mathbf{x}_{15}\mathbf{x}_{26}}{\mathbf{x}_{12}\mathbf{x}_{56}}|^2
+\bar{R}_{456}\log\ab|\frac{\mathbf{x}_{14}\mathbf{x}_{56}}{\mathbf{x}_{45}\mathbf{x}_{16}}|^2
+\bar{R}_{234}\bar{R}_{456}\log\ab|\frac{\mathbf{x}_{12}\mathbf{x}_{45}}{\mathbf{x}_{14}\mathbf{x}_{25}}|^2 \nonumber 
\\
&\quad +\bar{R}_{236}(\bar{R}_{456}-1)\log\ab|\frac{\mathbf{x}_{25}\mathbf{x}_{16}}{\mathbf{x}_{12}\mathbf{x}_{56}}|^2
+R_{345}\log\ab|\frac{\mathbf{x}_{13}\mathbf{x}_{56}}{\mathbf{x}_{15}\mathbf{x}_{36}}|^2
+\bar{R}_{456}R_{345}\log\ab|\frac{\mathbf{x}_{45}\mathbf{x}_{36}}{\mathbf{x}_{34}\mathbf{x}_{56}}|^2 \nonumber
\\
&\quad +\bar{R}_{234}\bar{R}_{456}R_{345}\log\ab|\frac{\mathbf{x}_{34}\mathbf{x}_{25}}{\mathbf{x}_{23}\mathbf{x}_{45}}|^2+\bar{R}_{236}R_{345}(\bar{R}_{456}-1)\log\ab|\frac{\mathbf{x}_{23}\mathbf{x}_{56}}{\mathbf{x}_{25}\mathbf{x}_{36}}|^2 \:.
\end{align}
\endgroup

\paragraph{Two loops and beyond.}  The contribution from the two-Reggeon exchange can be found in refs.~\cite{DelDuca:2016lad,Marzucca:2018ydt}. In the following, we focus on the three-Reggeon exchange and give more details for the two-loop NMHV case.

The Feynman diagrams are almost identical to those in eq.\ \eqref{eq: MHV_Three_Reggeon}, except that the vertices for the negative-helicity particles are replaced by the appropriate ones. For example, the Feynman integrals for the helicity configurations ${\ominus}{\oplus}{\oplus}{\oplus}$ and ${\oplus}{\ominus}{\oplus}{\oplus}$ are
\begin{equation}\begin{split}
    \mathcal{I}_{\{4,7\}}^{(2),\ominus\oplus\oplus\oplus} &= \int
    \frac{\mathbf{x}^{\ast}_{12} \mathbf{x}^{\ast}_{16} \mathbf{x}^{\ast}_{23} \mathbf{x}^{\ast}_{56} \mathbf{x}_{13} \mathbf{x}^{\ast}_{ab}\:\dbar^2 \mathbf{x}_{a}\dbar^2 \mathbf{x}_{b}}{\mathbf{x}^{\ast}_{13} |\mathbf{x}_{1a}|^2 \mathbf{x}^{\ast}_{a2} \mathbf{x}^{\ast}_{a6} \mathbf{x}^{\ast}_{2b} \mathbf{x}^{\ast}_{6b} \mathbf{x}^{\ast}_{b5} \mathbf{x}_{3b} \mathbf{x}_{ab}}  \:, \\
    \mathcal{I}_{\{4,7\}}^{(2),\oplus\ominus\oplus\oplus}&= \int\frac{\mathbf{x}^{\ast}_{13} \mathbf{x}^{\ast}_{16} \mathbf{x}^{\ast}_{56} \mathbf{x}_{12} \mathbf{x}_{14} \mathbf{x}_{23}
   \mathbf{x}^{\ast}_{b4}\:\dbar^2\mathbf{x}_{a}\dbar^2 \mathbf{x}_{b}}{\mathbf{x}^{\ast}_{14} \mathbf{x}_{13} |\mathbf{x}_{1a}|^2 \mathbf{x}^{\ast}_{a6}
   \mathbf{x}_{a2} \mathbf{x}^{\ast}_{3b} \mathbf{x}^{\ast}_{6b} \mathbf{x}^{\ast}_{b5} \mathbf{x}_{2b} \mathbf{x}_{b4}}  \:.
\end{split}\end{equation}
As in the two-loop MHV case, these two integrals are divergent due to the presence of the factor $1/|\mathbf{x}_{1a}|^2$. We adopt the same subtraction scheme as in eq.~\eqref{eq: subtraction_uni} and we find
\begingroup
\allowdisplaybreaks
\begin{align}
        \mathcal{I}_{\{4,7\}}^{(2),\ominus\oplus\oplus\oplus}\Big\vert_{\text{eq. \eqref{eq: subtraction_uni}}} &= \frac{\bar{R}_{235}}{2}\Bigl[
        \calG_{1}(\bar{\rho}_{1})(\calG_{1}(\bar{\rho}_{3})-\calG_{\bar{\rho}_{1}}(\bar{\rho}_{3}))
        -\calG_{0}(\bar{\rho}_{1})(\calG_{1}(\bar{\rho}_{3})+\calG_{\bar{\rho}_{1}}(\bar{\rho}_{3})) \nonumber  \\
        &\qquad\qquad-\calG_{1,\bar{\rho}_{1}}(\bar{\rho}_{3})+\calG_{\bar{\rho}_{1},1}(\bar{\rho}_{3})
     \Bigr] \nonumber \\
     &\quad +\frac{\bar{R}_{236}}{2}\Bigl[
        \calG_{0}(\bar{\rho}_1)(\calG_{1}(\bar{\rho}_3)+2\calG_{\bar{\rho}_{1}}(\bar{\rho}_{3})) 
        -\calG_{0,1}(\bar{\rho}_{1})+\calG_{1,0}(\bar{\rho}_{1}) \nonumber \\
        &\hspace{5em} -2\calG_{0,1}(\bar{\rho}_{3}) +2\calG_{0,\bar{\rho}_1}(\bar{\rho}_{3})
        \Bigr] \:, \\
             \mathcal{I}_{\{4,7\}}^{(2),\oplus\ominus\oplus\oplus}\Big\vert_{\text{eq. \eqref{eq: subtraction_uni}}} &= 
             -\frac{R_{234}}{2}\log\ab|1-\frac{\rho_{2}}{\rho_{3}}|^2 \log\ab|\frac{\rho_{1}}{(1-\rho_1)(1-\rho_{3})}|^2 \nonumber \\
            &\quad  +\frac{R_{234}\bar{R}_{345}}{2} \log\ab|\frac{\rho_{1}(\rho_{2}-\rho_{3})}{(\rho_{1}-\rho_{2})\rho_{3}}|^2\log\ab|\frac{\rho_{1}}{(1-\rho_1)(1-\rho_{3})}|^2 \nonumber \\
             &\quad  +\frac{R_{234}\bar{R}_{346}}{2} \log\ab|1-\frac{\rho_{2}}{\rho_{1}}|^2\log\ab|\frac{\rho_{1}}{(1-\rho_1)(1-\rho_{3})}|^2 \:.
\end{align}
\endgroup
We have verified that these two subtracted results agree with the multi-Regge limit of the two-loop NMHV octagon computed in ref.~\cite{He:2019jee}.

The three-Reggeon exchange for the other helicity configurations at two loops and for non-MHV amplitudes at higher loop orders can be obtained in a similar way. Their expressions are too lengthy to be recorded here, and the results up to three loops are collected in the file \texttt{nonMHV3Reggeon.m}.

\section{Conclusion}
\label{sec:conclusion}

In this paper, we have presented, for the first time, results for eight-point amplitudes in MRK at NLLA in the zigzag region, where both two- and three-Reggeon exchanges must be taken into account. Our main result consists of two ingredients. First, three-Reggeon exchange contributes at LLA, and its contribution can be evaluated explicitly using the BFKL-EFT. We have presented a simple set of Feynman rules that, after a suitable subtraction of IR divergences, allows the relevant Feynman diagrams to be readily evaluated in terms of single-valued polylogarithms using techniques from complex analysis. Second, we have proposed a representation of the two-Reggeon exchange contribution in terms of a compact Fourier-Mellin integral. Its integrand involves the same ingredients as those appearing in previously studied regions. In particular, these ingredients are known to all orders from integrability. The only difference from the results in other regions lies in the choice of integration contour for the Fourier-Mellin integral.

As a check of our proposal, we have shown that it reproduces the expressions for the multi-Regge limit in the zigzag region for the eight-point MHV amplitude up to three loops from refs.~\cite{Caron-Huot:2011zgw,Bargheer:2015djt,DelDuca:2018raq,He:2019jee,Li:2021bwg}. We have also presented novel results for the four-loop MHV and three-loop non-MHV octagons in MRK, as well as for the three-Reggeon contribution to higher-point MHV amplitudes in MRK. Our results provide novel insights into the analytic structure of higher-point scattering amplitudes, and we expect them to aid the future determination of these amplitudes in full kinematics, for example by providing boundary conditions for a bootstrap program for higher-point amplitudes.

A natural next step is to resum the large logarithms associated with three-Reggeon exchange. The corresponding rapidity evolution is described by the noncompact $\mathrm{SL}(2,\mathbb{C})$ spin chain introduced by Lipatov \cite{Lipatov:1993yb,Faddeev:1994zg,Lipatov:2009nt}. Three-Reggeon exchange represents the first genuinely nontrivial instance of this spin chain: while the two-Reggeon evolution involves only a single two-site Hamiltonian, the three-Reggeon evolution involves overlapping
two-site Hamiltonians whose interplay entangles all three Reggeon degrees of
freedom. It would be particularly interesting to clarify how this spin-chain integrability is related to the other integrable structures of planar $\mathcal{N}=4$ SYM~\cite{Beisert:2010jr} and whether it persists beyond LLA. Ultimately, one may hope that, in analogy with the two-Reggeon case \cite{Basso:2014pla,DelDuca:2019tur}, integrability can be used to determine all the building blocks required for a resummation of the three-Reggeon contribution to all orders.

The methods developed here may also have applications beyond planar $\mathcal{N}=4$ SYM. In QCD, the exchange of three reggeized gluons in a color-singlet, charge-conjugation-odd state gives rise to the \emph{Odderon} \cite{Lukaszuk:1973nt,Ewerz:2005rg}. Although the color representation and dynamical setting considered here differ from those relevant to the physical QCD Odderon, a better understanding of the three-Reggeon system, its analytic structure, and its integrability may provide useful insights and computational tools for the study of Odderon exchange and related high-energy phenomenology.

\acknowledgments
We are indebted to Xuhang Jiang for bringing the errors in the original ancillary files to our attention.
The work of CD and CZ is supported in part by the European Union (ERC Consolidator Grant LoCoMotive
101043686) and by the Deutsche Forschungsgemeinschaft (DFG, German Research Foundation) under Germany’s Excellence Strategy – EXC 3107 – Project-ID 53376636. 
The work of ZL was supported in part by the U.S. Department of Energy (DOE) under contract
number DE-AC02-76SF00515.
Views and opinions expressed are however those of the
author(s) only and do not necessarily reflect those of the European Union or the European
Research Council. Neither the European Union nor the granting authority can be held
responsible for them. 

\appendix

\section{The multi-Regge limit of the three-loop MHV octagon} \label{app:A}

In this appendix, we shall follow ref.~\cite{DelDuca:2018raq} and briefly review how to analytically continue the octagon remainder function to the Mandelstam region $\varrho$ at the level of its symbol.

We start with the known result for the three-loop octagon~\cite{Li:2021bwg}, whose symbol can be organized as
\begin{align} \label{oct3loop_symbol1}
    \Delta_{2,1,1,1,1}R_{8}^{(3)} &= \sum_{i\leq k} (\log U_{ij}\log U_{kl}) \otimes A_{ijkl} -
    \sum_{ijkl}\Li_{2}(1-U_{ijkl})\otimes B_{ijkl} \nonumber \\
    &\quad + I_{\text{box}}^{(1,3,5,7)}\otimes A^{(1)} + I_{\text{box}}^{(2,4,6,8)}\otimes A^{(2)} \:,
\end{align}
where we have introduced cross-ratios
\begin{equation}
   U_{ijkl}:= \frac{x_{ij}^2 x_{kl}^2}{x_{ik}^2 x_{jl}^2}, \quad U_{ij}:=U_{i,j+1,j,i+1} \:,
\end{equation}
and we require the subscripts of $U_{ij}$ to satisfy $i<j$, since $U_{ij}=U_{ji}$. The arguments of $\Li_{2}$ are choosen such that they become 0 or 1 in the multi-Regge limit, which are 
\begin{align*}
    &U_{14} \text{ and 7 cyclic images}, \quad U_{15} \text{ and 3 cyclic images},\\
    &U_{14}U_{15} \text{ and 7 cyclic images}, \\
    &U_{15}U_{16} \text{ and 7 cyclic images}, \\
    &U_{14}U_{15}U_{16} \text{ and 7 cyclic images}.
\end{align*}
Note that there are 36 $\Li_{2}$ functions in total and this set is invariant under permutations and reflections.  A new ingredient that appears for the first time in the octagon at three loops is the four-mass box function,
\begin{equation} \label{eq:box-function}
    I_{\text{box}}^{(i,j,k,l)}=- \Li_{2}(z)+\Li_{2}(\bar{z})-\frac{1}{2}\log(z\bar{z})\log\frac{1-z}{1-\bar{z}}\,,
\end{equation}
where 
\begin{equation}
    z\bar{z}=U_{ijkl} \:,\qquad (1-z)(1-\bar{z})=U_{jkli} \:.
\end{equation}

\subsection{Multi-Regge limit in the Euclidean region}

BDS-normalized amplitudes vanish in multi-Regge limits in the Euclidean region, where all $x_{ij}^{2}<0$. This imposes several constraints on the multi-Regge limits of $A_{ijkl}$, $B_{ijkl}$ and $A^{(1,2)}$ in eq.~\eqref{oct3loop_symbol1}. First, note that 
\begin{equation}\begin{split}
    \calM(I_{\text{box}}^{(1357)})&=\calM\!\!\left[-\Li_{2}(1-U_{3571})-\frac{1}{2}\log U_{1357}\log U_{3571}\right]\:,  \label{eq:MRK_box(a)}\\
    \calM(I_{\text{box}}^{(2468)})&=\calM\!\!\left[-\Li_{2}(1-U_{4682})-\frac{1}{2}\log U_{2468}\log U_{4682}\right] \:.
\end{split}
\end{equation}
In the Euclidean region, $\calM(I_{\text{box}}^{(2468)})$ simply becomes zero (due to $\calM(U_{4682})=1$) and therefore puts no constraint on $A^{(2)}$. However, $\calM(I_{\text{box}}^{(1357)})$ becomes divergent, due to the term $\log U_{1357}\log U_{3571}$, which has the same divergence as $(\log U_{38} + \log U_{48})(\log U_{15} + \log U_{16})$. This requires
\begin{equation}
    \frac{1}{2}\calM(A^{(1)})=\calM(A_{1538})=\calM(A_{1548})=\calM(A_{1638})=\calM(A_{1648})\:,
\end{equation} 
while $\calM(A_{1ij8})=0$ otherwise. Furthermore, $\calM(A_{1j1l})=\calM(A_{i8k8})=0$ due to the same reason as at two loops.
These constraints should hold to \emph{all loop orders}. At three loops, we find in addition the following relations:
\begin{equation}\begin{split}
    \frac{1}{2}\calM(A^{(1)})&=\calM(A_{1538})=\calM(A_{1548})=\calM(A_{1638})=\calM(A_{1648})  \\
    &=\calM(A_{1547})=\calM(A_{2547})=\calM(A_{2548})\,, \\
    \frac{1}{2}\calM(A^{(2)})&=\calM(A_{1436})=\calM(A_{1437})=\calM(A_{2658})=\calM(A_{3658})\,, \\
   0&= \calM(A_{1525})=\calM(A_{1625})=\calM(A_{1626})=\calM(A_{1636})=\calM(A_{3638})  \\
   &= \calM(A_{3738})=\calM(A_{1447})=\calM(A_{3847})=\calM(A_{4748})=\calM(A_{2558})\,.
\end{split}\end{equation}
It is unclear whether these relations are merely coincidental or have deeper underlying reasons.

\subsection{Mandelstam regions and analytic continuations of first two entries}

To determine the multi-Regge limit of the BDS-normalized amplitudes $R_{N}^{h_1\cdots h_N}$, we need to know the behavior of the first entries of $\Delta_{2,1,\ldots,1} R_{N}^{h_1\cdots h_N}$ in different Mandelstam regions. When they are $\log U_{ij}\log U_{kl}$ and $\Li_{2}(1-U_{ijkl})$, their behavior is relatively simple and given by
\begin{align}
	\begin{split}
	\log U_{ij}\,=\,&\log\ab|U_{ij}| + 2\pi i\, n^{\varrho}_{ij}\ ,\\[5pt]
	\label{eq:Li2ac}
	\Li_2\left(1-U_{ijkl}\right) \,=\,&  \text{Re}\,\text{Li}_2\left(1-U_{ijkl}\right) -2\pi i\, n_{ijkl}^{\varrho} \log\ab|1-U_{ijkl}|\\
	&-\frac{1}{2}\,(2\pi i)^2 (n^\varrho_{ijkl})^2\,\theta\left(U_{ijkl}-1\right),
	\end{split}
\end{align}
where $n_{ij}^{\varrho}$ and $n_{ijkl}^{\varrho}$ are the winding numbers of $U_{ij}$ and $U_{ijkl}$ in region $\varrho$, respectively. Explicitly, we have~\cite{Bargheer:2015djt}
\begin{align}
  n^{\varrho}_{ijkl} &= -\sum_{a=i}^{l-1}\sum_{b=j}^{k-1} n_{ab}^{\varrho}, \quad \text{for } i<j<k\text{ and } i<l, \\
  n^{\varrho}_{ij} &=\begin{cases}
    -\frac{1}{4}(\varrho_{j}-\varrho_{j+1}), \quad  &\text{if } i=1, \\
     \phantom{-}\frac{1}{4}(\varrho_{i+1}-\varrho_{i+2}), \quad &\text{if } j=N, \\
     -\frac{1}{4}(\varrho_{i+1}-\varrho_{i+2})(\varrho_{j+1}-\varrho_{j})\:, \quad &\text{otherwise}.
  \end{cases}
\end{align} 

We are concerned with the BDS-normalized three-loop MHV amplitude $R_{8}^{(3)}$, and for $\Delta_{2,1,\ldots,1}R_{8}^{(3)}$ the first entries also include the four-mass box function in eq.~\eqref{eq:box-function}, whose analytic continuation to other kinematic regions is more complicated (see ref.~\cite{Corcoran:2020akn} for a recent discussion of this problem). However, since we are only interested in the multi-Regge limit, we can simply take the multi-Regge limit for the four-mass boxes in the Euclidean region and then analytically continue them to other regions. In other words, in MRK, we have
\begin{equation}
I_{\text{box}}^{(ijkl)} \simeq {-}\Li_{2}(1-U_{jkli})-\tfrac{1}{2}\log U_{ijkl}\log U_{jkli}\,.
\end{equation}

\section{The Two-Reggeon Exchange Contribution \texorpdfstring{$\mathcal{W}_{\{4,7\}}$}{W\_47}} \label{app:B}

In this section, we shall briefly describe how to obtain $\mathcal{W}_{\{4,7\}}$ from the known result $\mathcal{W}_{[k,l]}$ through eq.~\eqref{eq: nonMHV_two_reggeon_sub} by using the known perturbative results available in the literature, in particular refs.~\cite{Marzucca:2018ydt,DelDuca:2019tur}.

Schematically, the general two-Reggeon-exchange contribution $\mathcal{W}_{\varrho,N}^{\vec{h}}$, where $\vec{h}$ denotes the helicity sequence $h_{4}\ldots h_{N-1}$, can be written as
\begin{align}
  \mathcal{W}_{\varrho,N}^{\vec{h}}=\sum_{L=1}^{\infty} (2g^2)^{L}
  &\left[\sum_{|\vec{i}|\leq L-1}\ab(\prod_{a=1}^{N-5}\frac{1}{i_{a}!}\log^{i_{a}}\tau_{a}) \tilde{g}_{\varrho,\vec{h}}^{(L;\vec{i}\,)}(z_{1},\ldots,z_{N-5})\right.  \nonumber \\
  &\left.\quad +(2\pi\mi)\sum_{|\vec{i}|\leq L-2}\ab(\prod_{a=1}^{N-5}\frac{1}{i_{a}!}\log^{i_{a}}\tau_{a}) \tilde{h}_{\varrho,\vec{h}}^{(L;\vec{i}\,)}(z_{1},\ldots,z_{N-5})\right] \:,
\end{align}
where $\vec{i}$ denotes the sequence of nonnegative integers $i_{1},\ldots,i_{N-5}$, and $|\vec{i}|$ denotes their sum. Then, according to eq.~\eqref{eq: nonMHV_two_reggeon_sub}, the quantities of interest, $\tilde{h}_{\{4,7\},\vec{h}}^{(L;\vec{i})}$ with $|\vec{i}|=L-2$, can be expressed in terms of $\tilde{g}_{[k,l],\vec{h}}^{(L;\vec{i})}$ and $\tilde{h}_{[k,l],\vec{h}}^{(L;\vec{i})}$ as follows:
\begin{align}
    \tilde{h}_{\{4,7\},\vec{h}}^{(L;\vec{i})}\Big|_{|\vec{i}|=L-2}
    &=
    \tilde{h}_{[4,7],\vec{h}}^{(L;\vec{i})}
    -\tilde{h}_{[4,6],\vec{h}}^{(L;\vec{i})}
    -\tilde{h}_{[5,7],\vec{h}}^{(L;\vec{i})}
    +\tilde{h}_{[5,6],\vec{h}}^{(L;\vec{i})}
    \nonumber\\
    &\quad
    -\frac{1}{2}\Bigl[
        \tilde{g}_{[4,7],\vec{h}}^{(L;\vec{i}+\vec{e}_{1})}
       +\tilde{g}_{[4,7],\vec{h}}^{(L;\vec{i}+\vec{e}_{3})}
       -\tilde{g}_{[4,6],\vec{h}}^{(L;\vec{i}+\vec{e}_{1})}
       -\tilde{g}_{[5,7],\vec{h}}^{(L;\vec{i}+\vec{e}_{3})}
    \Bigr]
    \nonumber\\
    &\quad
    +\frac{1}{4}\Bigl[
        \log |z_{3}|^2
        \tilde{g}_{[4,6],\vec{h}}^{(L-1;\vec{i})}
       -\log |z_{1}|^2
        \tilde{g}_{[5,7],\vec{h}}^{(L-1;\vec{i})}
       + \log|z_{1}/z_{3}|^2
        \tilde{g}_{[5,6],\vec{h}}^{(L-1;\vec{i})}
    \Bigr] \nonumber \\
 &\quad -\frac{1}{16}
\delta_{L,2} \log|z_{1}|^2 \log |z_{3}|^2 
\:,
    \label{eq:h_zigzag_explicit}
\end{align}
where
\begin{align}
    \vec{e}_{1}:=(1,0,0)\:,\qquad
    \vec{e}_{3}:=(0,0,1)\:.
\end{align}
The terms on the second line of eq.~\eqref{eq:h_zigzag_explicit} arise from the replacements $\log\tau_{1,3}\to\log\tau_{1,3}-\mi\pi$, while those on the last two lines arise from expanding the three additional phase factors in eq.~\eqref{eq: nonMHV_two_reggeon_sub}.

Explicit results for $g_{[4,7],\vec{h}}^{(L;\vec{i})}$ and $h_{[4,7],\vec{h}}^{(L;\vec{i})}$ through $L=3$ can be obtained from ref.~\cite{Marzucca:2018ydt}. The remaining cases follow simply by taking appropriate soft limits. At higher loops, these ingredients can be derived from the all-order Fourier-Mellin representation of ref.~\cite{DelDuca:2019tur}.

\bibliographystyle{JHEP}
\bibliography{refs}

\end{document}